\documentclass[12pt]{iopart}
\usepackage{graphicx}
\begin{document}

\title[Min oscillations and segregation during cell division]{A
stochastic model of Min oscillations in \textit{Escherichia coli} and Min
protein segregation during cell division} 

\author{Filipe Tostevin and Martin Howard}

\address{Department of Mathematics, Imperial College London, London
SW7 2AZ, UK}
\ead{filipe.tostevin@imperial.ac.uk}

\begin{abstract}
The Min system in \textit{Escherichia coli} directs division to the centre
of the cell through pole-to-pole oscillations of the MinCDE proteins. We
present a one dimensional stochastic model of these oscillations which incorporates
membrane polymerisation of MinD into linear chains. This model reproduces
much of the observed phenomenology of the Min system, including pole-to-pole
oscillations of the Min proteins. We then apply this model to investigate
the Min system during cell division. Oscillations continue initially unaffected
by the closing septum, before cutting off rapidly. The fractions of Min proteins
in the daughter cells vary widely, from 50\%-50\% up to 85\%-15\% of the
total from the parent cell, suggesting that there may be another mechanism
for regulating these levels \textit{in vivo}.
\end{abstract}
\submitto{\it Phys. Biol.\/}

\maketitle


\section{Introduction}

Reproduction of the rod-shaped bacterium \textit{Escherichia coli} takes
place through binary fission. In order for the production of viable daughter
cells the site of division must be accurately located at the middle of the
cell to ensure one chromosome is distributed to each daughter cell. This
positional information is provided by the combined action of nucleoid occlusion,
which inhibits division close to the chromosomes, and the proteins of the
Min system. Investigation of the dynamics of the Min proteins has decisively
shown that mathematical modelling and ideas from pattern-formation can be
useful in the study of subcellular processes \cite{HK}.

Cell division occurs
after the chromosome has been replicated and the daughter chromosomes
have moved into opposite halves of the cell. Nucleoid occlusion limits
division to regions without DNA \cite{Yu}, i.e. the cell's centre or 
poles. The Min system consists of the proteins MinC, MinD and
MinE. Its function is to prevent septal ring formation at the cell
poles, so the only favourable division site is at the cell centre,
equidistant from the two cell poles.

In the presence of MinE, the MinC and MinD proteins oscillate between
the two cell poles \cite{RdB,Hu1}. MinC and MinD are observed to
occupy the membrane in one half of the cell. A MinE ring forms at the
medial end of this region and moves towards the cell pole
\cite{Fu,Hale}, displacing the MinC and MinD in the process. The MinC
and MinD then gather in the opposite half of the cell, and the
process repeats. The period of these oscillations \textit{in vivo} is
30-120 seconds.

Several studies have been made with different
reaction-diffusion models to explain these oscillations
\cite{Meinhardt,Howard1,Kruse1,Howard2,Huang}. Subsequent experiments have
investigated the reaction steps involved in the Min oscillation cycle,
and have allowed the details of the models to be refined: MinD:ATP first
binds to the cell membrane. In the absence of MinE, MinD is
distributed evenly throughout the membrane \cite{RdB}. The rate of
MinD accumulation, through cooperative binding or self-aggregation, 
increases with the amount of MinD present \cite{lackner}. MinD
forms oligomers \cite{Hu2002,suefuji}, and can form a complex with
either MinC or MinE \cite{Huang96}. MinC inhibits polymerization of
FtsZ \cite{ZonglinHu12211999}, preventing formation of the ``Z-ring"
which forms the basis for the division machinery. MinD enhances the
effect of MinC by recruiting it to the membrane. MinC is co-localized 
with membrane-bound MinD
\cite{Hu1,RdB2}. However, MinC is not required for the oscillation of MinD
and MinE \cite{RdB}. MinE is recruited to the membrane by MinD where
it forms a MinDE complex and, in the process, expels MinC from the
membrane \cite{lackner}. MinE also stimulates ATP-hydrolysis of
membrane-bound MinD, which causes dissociation of MinD from the
membrane. MinD:ADP then undergoes nucleotide exchange in the cytoplasm
to MinD:ATP.

It has also recently emerged that MinD forms helical filaments in living
cells \cite{coils}; recent mathematical models \cite{Drew,Kruse2,Pavin} have
attempted to include this feature. In the Meacci-Kruse model \cite{Kruse2}
the membrane occupancy is limited, and MinD accumulation is due to self-aggregation
once it has bound to the membrane. The model by Drew \etal \cite{Drew} includes
polymer growth from nucleation sites at the ends of the cell, but this and
other assumptions upon which it relies, such as regulating polymer growth
rate according to length, are not required in other models. Both of these
models use continuous partial differential equations. The model by Pavin
\etal \cite{Pavin} differs in that it is a three-dimensional stochastic model,
but it does not form the observed large scale helical filaments. Incorporating
stochasticity (first introduced into Min modelling in \cite{Howard2}) is
nevertheless likely to be important for systems of this type: for example,
fluctuations are believed to be vital for modelling the closely related Spo0J/Soj
dynamics in {\it Bacillus subtilis} \cite{Doubrovinski}.

In this paper, we present a simple one-dimensional stochastic model that
reproduces many of the experimental observations of the Min oscillations.
Aside from including full fluctuation effects, our stochastic model tracks
individual protein particles, and thus makes it easier to, (i), incorporate
the structure of the Min proteins on the membrane, (ii), limit the maximum
membrane occupancy, and, (iii), allow binding rates to depend on the local
arrangement of the proteins. Importantly, we allow the MinD to form linear
membrane-bound polymers along the cell length. However, as we will see, oscillatory
dynamics can be reproduced independent of many of the details of the polymer
structure. In our model we have therefore chosen a particularly simple implementation
of membrane polymerisation. We also assume that proteins incorporated into
membrane-bound polymers are fixed in place and cannot diffuse. This difference
in mobility between the membrane and cytoplasm is crucial for enabling pattern
formation in our model.

Although the Min oscillations have been studied in detail, there have only
been a few comments describing oscillations in constricting and recently
divided cells \cite{RdB,Hu1}. We therefore use our model to investigate the
Min system during these phases by incorporating division at the centre of
the cell into the simulations. We find that the dynamics of the Min proteins
during contraction of the Z-ring is generally consistent with the available
experimental observations: the pole-to-pole oscillations continue for some
time and then the dynamics changes sharply to independent oscillations on
each side of the septum. We also study the numbers of Min proteins that are
found in the two daughter cells. The numbers of Min proteins in each half
of the parent cell vary greatly over the pole-to-pole oscillation period,
and we find that the protein numbers in the daughter cells also vary from
cell to cell over a similar range. This result suggests that the number of
Min proteins may fluctuate strongly from cell to cell, but also that there
may be other mechanisms for controlling protein numbers \textit{in vivo},
such as the rates of Min protein synthesis being regulated by the Min protein
concentration levels.


\section{The Model}
\label{model}
The cell is modelled in 1-dimension by dividing the length $L$
into $N$ discrete intervals of width $dx=L/N$. Each interval $i$ contains
$n_p^i$ of each of the five protein states in the model. These are
cytoplasmic MinD:ADP ($ p={\rm D:ADP}$), cytoplasmic MinD:ATP ($p={\rm D:ATP}$),
cytoplasmic MinE ($p={\rm E}$), membrane-bound MinD ($p={\rm d}$), and
membrane-bound MinDE complex ($p={\rm de}$). MinE is present as a homodimer
\cite{king1999}, so one MinE unit is actually a dimer rather than a
single protein. Experiments show that MinC is not required for the
oscillations, so it is not included explicitly in the modelling. Since
MinC is co-localized with MinD in a MinCD complex, we assume that the
amount of membrane-bound MinC can be quantified by measuring
$n_{\rm d}^i$. In our simulations we use a time step $dt$. Simulations begin
with either uniform initial protein distributions or random distributions
without affecting our results.

\begin{figure} 
\centering 
 \includegraphics[width=0.6\textwidth]{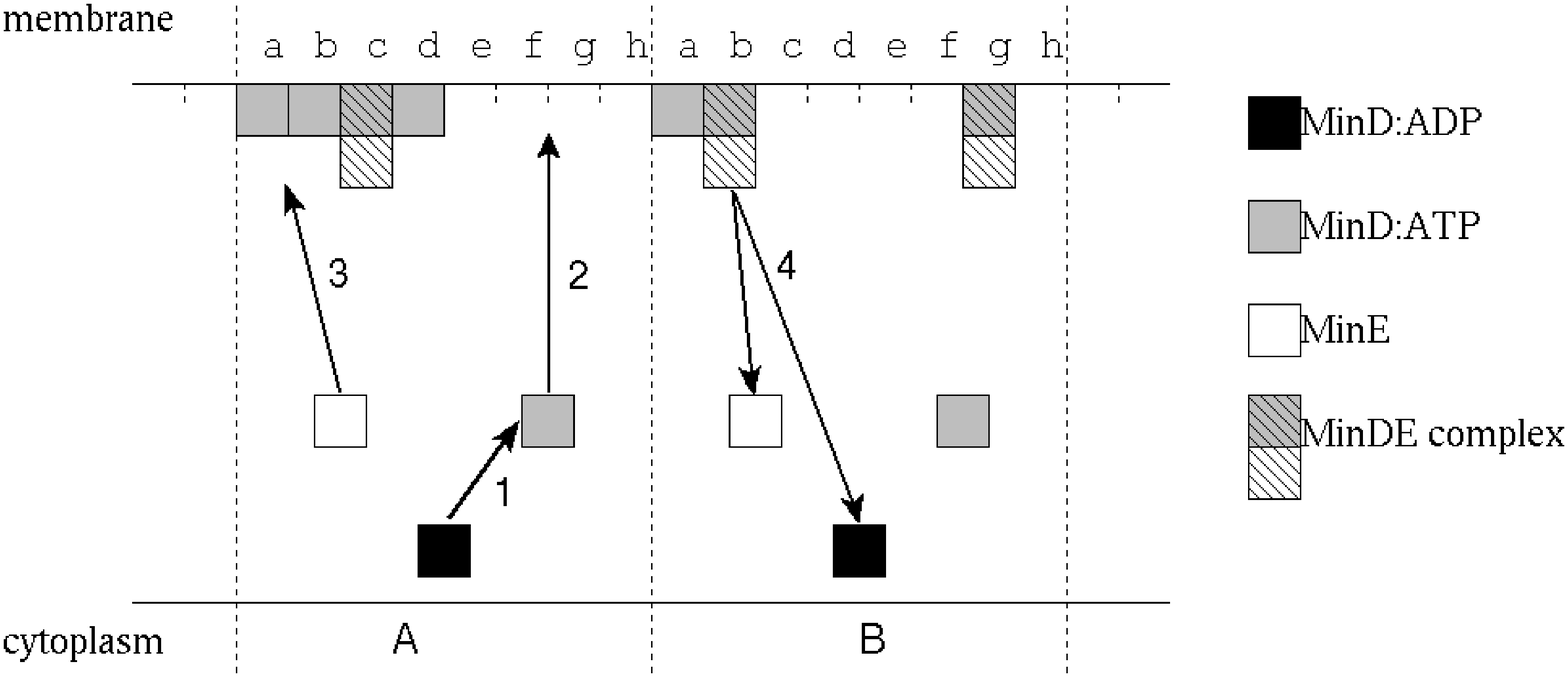} 
\caption{Schematic showing the model steps with one membrane filament ($N_{\rm
c}=1$), with $n_{\rm max}=8$: 1. MinD:ADP converts to MinD:ATP. 2. MinD:ATP
binds to the membrane.  In this case, a MinD:ATP in cytoplasmic site A could
bind at membrane position Ae or Ah with probability $\sigma_{\rm d,coop}dt$,
or at Af or Ag with probability $\sigma_{\rm d,sp}dt$. MinD:ATP in cytoplasmic
site B could bind with the lower probability to each empty site as there
are no suitable sites for cooperative binding. 3. MinE binds to membrane
MinD with probability $\sigma_{\rm e}dt$ per binding site. 4. The MinDE complex
dissociates, giving cytoplasmic MinD:ADP and MinE. The complex would unbind
from site Bg with probability $\sigma_{\rm dis,iso}dt$, since both neighbouring
sites are empty; from Bb with probability $\sigma_{\rm dis,end}dt$; and from
Ac with probability $\sigma_{\rm dis,bulk}dt$, since both neighbouring sites
are occupied.}
\label{mech} 
\end{figure}

Membrane filaments are modelled by subdividing the cell membrane into
$N_{\rm c}$ linear arrays of $n_{\rm max}$ possible binding sites for each
of the $N$ discrete intervals. Each of the $N_{\rm c}$ arrays extends along
the length of the cell, allowing filaments to grow regardless of the
discretization boundaries in the cytoplasm. During the reaction steps,
cytoplasmic molecules may bind to membrane sites contained within the
interval they currently occupy, as shown in figure \ref{mech}. Each of
these membrane sites influences only its immediate neighbours on the
membrane, and any molecules occupying neighbouring sites are
considered bound in a polymer chain. The dynamical behaviour is
independent of the values of $N_{\rm c}$ and $n_{\rm max}$ provided the 
total number of membrane binding sites per cytoplasmic site, 
$N_{\rm c}n_{\rm max}$, is maintained. This result suggests that the overall
number of MinD molecules that can bind to the membrane influences the dynamical
behaviour, but the number of filaments into which they are arranged does
not.

In this paper we employ a particularly simple way to incorporate polymerisation,
with a minimal number of assumptions about the {\it in vivo} polymerisation
and structure. Since we can reproduce the Min oscillations with this model,
the exact details of polymerisation appear not to be important for generating
the experimentally observed Min dynamics. In particular we include only the
basic effects which any more advanced polymerisation model must also contain,
the most important of which is reduced mobility for proteins which are membrane
bound. In the model, once bound to the membrane a molecule cannot
move and is fixed in place until it dissociates from the membrane. We have
also tested the model with diffusion of isolated membrane MinD with a similar
diffusion constant to that in the cytoplasm. This change has no effect on
the behaviour of the model as the amount of isolated membrane MinD is small
compared to the amount of membrane MinD bound together into polymers.

All cytoplasmic proteins diffuse with diffusion constant $D$. The
probability of a molecule moving to the left or right,
\begin{equation}
 n^i_p \rightarrow n^i_p-1, \  n^{i\pm 1}_p \rightarrow n^{i\pm 1}_p+1,
\end{equation}
in a time interval $dt$ is $D dt/{(dx)}^2$.

MinD:ATP binds to the cell membrane:
\begin{equation}
 n^i_{\rm D:ATP} \rightarrow n^i_{\rm D:ATP}-1, \  n^i_{\rm d} \rightarrow n^i_{\rm d}+1.
\end{equation}
Cooperative binding and self-assembly of MinD are simulated by using
two different rates for membrane attachment. If a MinD molecule is
present on the membrane and a neighbouring membrane site is empty,
cytoplasmic MinD:ATP will bind with probability $\sigma_{\rm d,coop}dt$
for each such site. MinD may also bind to any other empty site with a
much lower probability, $\sigma_{\rm d,sp}dt$. Since the binding rate is
much higher if there is already MinD on the membrane, polymer chains
form as protein particles preferentially bind to the MinD already
present. In the model, MinD is not allowed to bind cooperatively to
the MinDE complex. If this reaction is allowed to take place at the
faster rate $\sigma_{\rm d,coop}$, then oscillations do not occur. MinD is
allowed to bind adjacent to the MinDE complex, but at the slower rate
$\sigma_{\rm d,sp}$. We consider that MinE at the end of a polymer blocks
the tendency for self-assembly, but cannot completely block MinD
binding.

Cytoplasmic MinE may bind to a membrane-bound MinD molecule, with
probability $\sigma_{\rm e}dt$ for each such site, forming the MinDE
complex:
\begin{equation}
 n^i_{\rm E} \rightarrow n^i_{\rm E}-1, \ n^i_{\rm d} \rightarrow n^i_{\rm d}-1, \
 n^i_{\rm de} \rightarrow n^i_{\rm de}+1.
\end{equation}

Dissociation of the complex releases one MinD:ADP molecule and one
MinE dimer into the cytoplasm:
\begin{equation}
 n^i_{\rm de} \rightarrow n^i_{\rm de}-1, n^i_{\rm D:ADP} \rightarrow
 n^i_{\rm D:ADP}+1, \ n^i_{\rm E} \rightarrow n^i_{\rm E}+1.
\end{equation}
There are three rates for dissociation, depending on the position in the
membrane array. The fastest rate and hence highest probability, $\sigma_{\rm
dis,iso}dt$, is for isolated molecules of the MinDE complex, which have no
immediate neighbours on the membrane. The complex unbinds from the end of
a chain (i.e. if it has one empty neighbouring site) with lower probability
$\sigma_{\rm dis,end}dt$, and from within a chain (neither neighbouring site
empty) with a still lower probability $\sigma_{\rm dis,bulk}dt$. These slower
rates result from the existence of bonds to neighbouring units in the polymer
chain. However, these different rates are not required for the oscillations,
which can be achieved with a single dissociation rate independent of position.
This suggests that the cooperative binding and reduced mobility introduced
by polymerisation are more important in generating oscillations than the
details of disassembly. However, we still include these three rates to take
account of the polymer nature of the membrane proteins.

MinD is released from the membrane in the MinD:ADP form. Before it is
able to rebind it must undergo nucleotide exchange to the MinD:ATP
form:
\begin{equation}
 n^i_{\rm D:ADP} \rightarrow n^i_{\rm D:ADP}-1, \ n^i_{\rm D:ATP} \rightarrow
 n^i_{\rm D:ATP}+1.
\end{equation}
This occurs in an interval $dt$ with probability $\sigma_{\rm DT}dt$. This
reaction step is also not required for the oscillations, but its
inclusion makes the model more robust to changes in protein numbers.

\subsection{Parameters}

We use $dx=0.01\mu m$ and $dt=10^{-5}s$. We have checked that reducing $dt$
by a factor of 10, or reducing $dx$ by a factor of 4 while keeping $L$ and
the total number of membrane sites constant, does not affect our results.
We take $N_{\rm c}=2$ since observations suggest that there are about two
independent helical MinD filaments in living cells \cite{coils}. In our model
there is no interaction between different filaments, since they are likely
to be spaced far apart on the cell membrane. MinD proteins have a length
of approximately $5nm$ \cite{suefuji}. Assuming that during polymerisation
there is some overlap or interlocking, and that the helical filaments have
a relatively large angle with the cell's long axis \cite{coils}, we assume
it takes 6 MinD molecules to span the $dx=0.01\mu m$ interval. Furthermore,
MinD polymers are likely to be double-stranded \cite{suefuji}, and we have
therefore taken $n_{\rm max}=2\times 6=12$. However, we have observed oscillations
for $N_{\rm c}n_{\rm max}$ in the range 12-30 and $N_{\rm c}$ from 1 to 4,
indicating a high degree of robustness in the values of these parameters.
For smaller $N_{\rm c}n_{\rm max}$ values, MinD fails to form the high density
polar regions required for oscillation, instead filling the membrane uniformly.
For larger $N_{\rm c}n_{\rm max}$, large amounts of MinD are able to gather
in small regions, and as a result regions of high MinD concentration are
not observed to extend long distances across the cell.

Unless otherwise specified, simulations are performed with $L=3\mu
m$. The densities used are $\rho_{\rm D}=1000\mu m^{-1}$ MinD protein
particles and $\rho_{\rm E}=400\mu m^ {-1}$ MinE homodimers
\cite{shih2002}. We use $D=2.0\mu m^2s^{-1}$, from experimental
measurements of the diffusion rates of (unrelated) cytoplasmic proteins in
\textit{E.  coli} \cite{diff.}. The other parameters take the
following values: $\sigma_{\rm DT}=1s^{-1}$, $\sigma_{\rm d,sp}=0.005s^{-1}$,
$\sigma_{\rm d,coop}=30s^{-1}$, $\sigma_{\rm e}=50s^{-1}$,
$\sigma_{\rm dis,iso}=10s^{-1}$, $\sigma_{\rm dis,end}=0.3s^{-1}$, and
$\sigma_{\rm dis,bulk}=0.1s^{-1}$.

These values were chosen to fit the results of the model with experimental
results, particularly the oscillation period. Increasing $\sigma_{\rm DT}$
increases the period, since MinD is able to rebind more quickly and will
therefore rebind more times within one polar zone before diffusing to the
opposite pole of the cell. $\sigma_{\rm dis,end}$ controls the rate at which
MinD polar zones are disassembled, and hence also has a significant effect
on the period. However, the fundamental oscillatory dynamics are robust to
significant changes in each of the parameter values individually. For example,
oscillations persist if $\sigma_{\rm d,coop}$ or $\sigma_{\rm e}$ are changed
by a factor of 2. The values of $\sigma_{\rm dis,bulk}$, $\sigma_{\rm dis,iso}$
and $\sigma_{\rm d,sp}$ have little effect on the dynamics, as long as $\sigma_{\rm
d,sp}<<\sigma_{\rm d,coop}$, although increasing $\sigma_{\rm d,sp}$ or decreasing
$\sigma_{\rm dis,iso}$ does lead to increased noise in the oscillatory pattern.


\subsection{Results}

\textit{Pole-to-pole oscillations:}
Initially there is a transient period which lasts about one to two
minutes, during which pole-to-pole oscillations are established. After
this time, the oscillations are stable and persist over at least 90
minutes of simulated time.

\begin{figure}
 \includegraphics[width=0.46\textwidth]{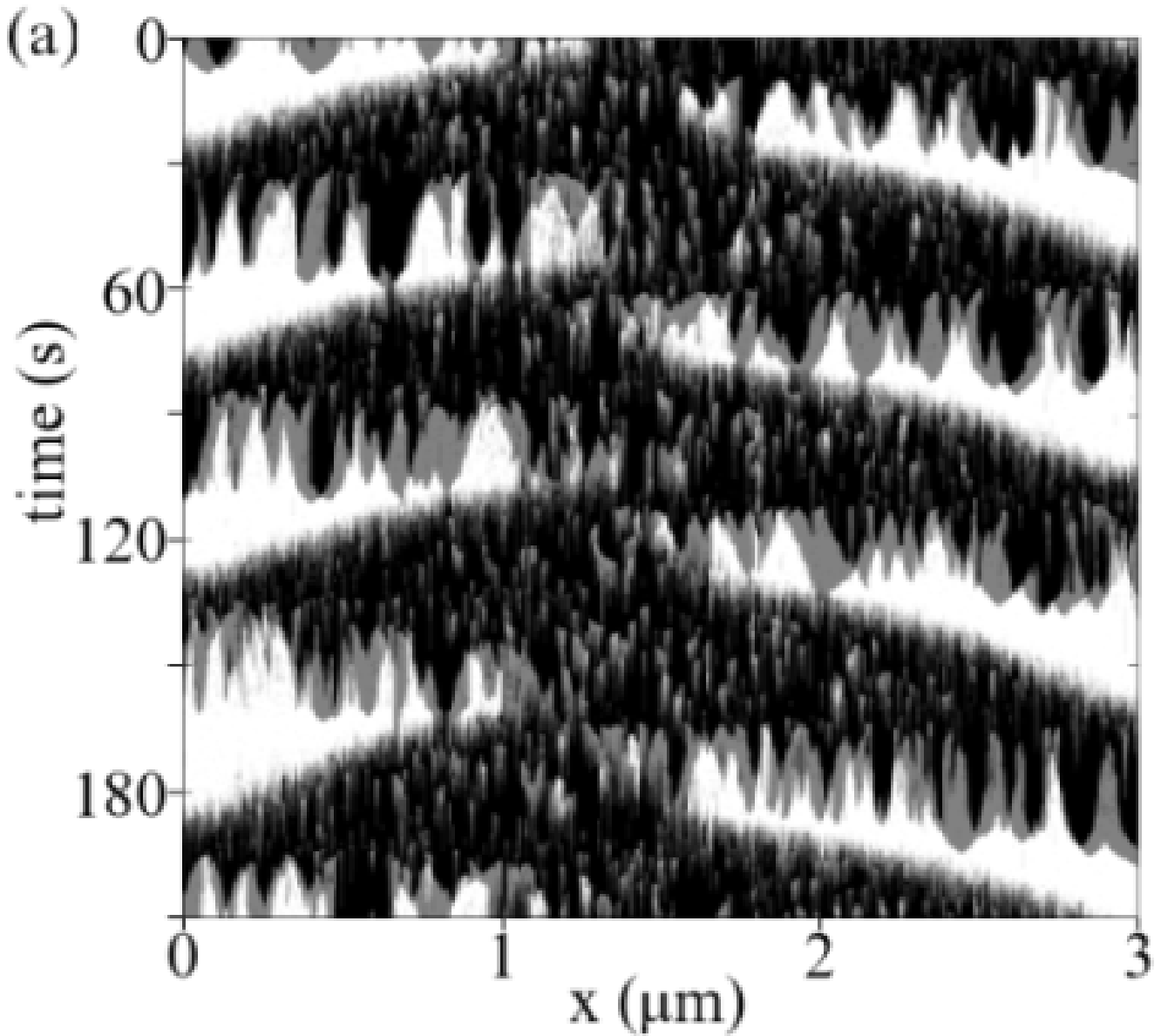} 
 \includegraphics[width=0.46\textwidth]{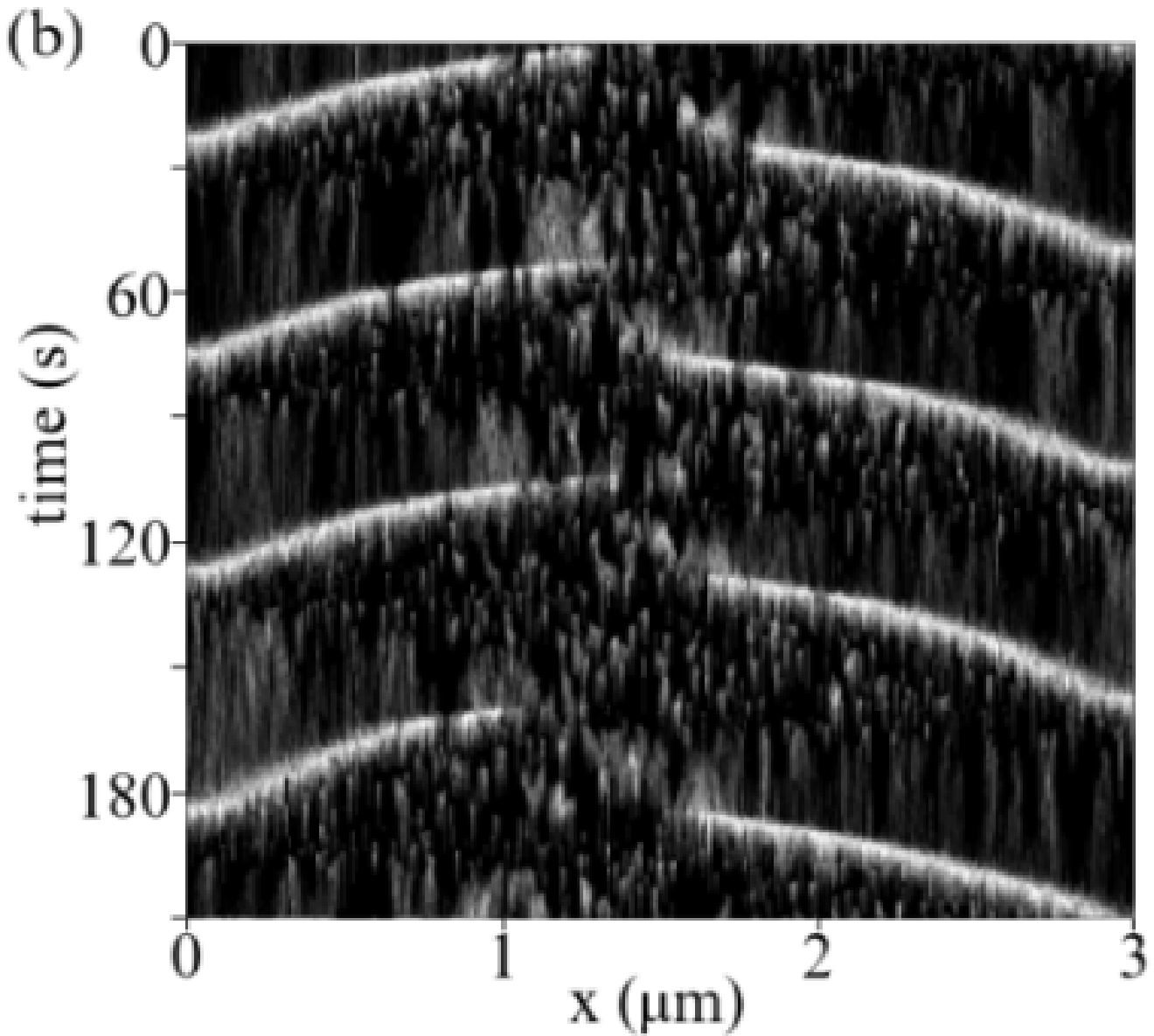}
 \includegraphics[width=0.063\textwidth]{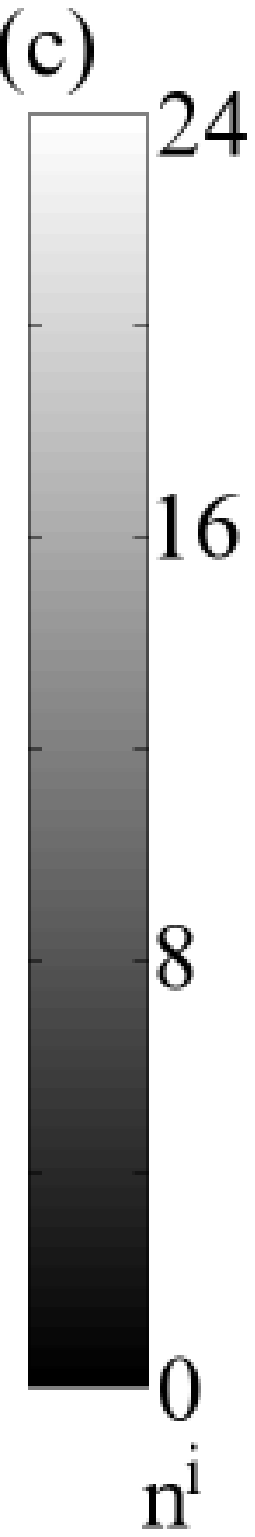} 
 \caption{Space-time plot of protein densities for ($a$)
 membrane-bound MinD, and ($b$) the MinDE complex; ($c$) 
 shows the scale used.}
 \label{standard}
\end{figure}

In our model, MinD filaments tend not to grow out from the cell poles, instead
the MinD filaments grow from random sites in the half of the cell where the
concentration of MinE is lower. This is in contrast with experiment, where
MinD polar regions often grow from the cell pole towards midcell. This difference
in behaviour is a general feature of our model, independent of specific parameter
values. In particular, it is difficult to prevent binding away from the cell
pole because the MinE levels are low and roughly constant over this region.
A more significant change to the model, such as adding favourable binding
sites near the cell poles, could perhaps overcome these difficulties.

\begin{figure} 
 \centering 
 \includegraphics[width=0.5\textwidth]{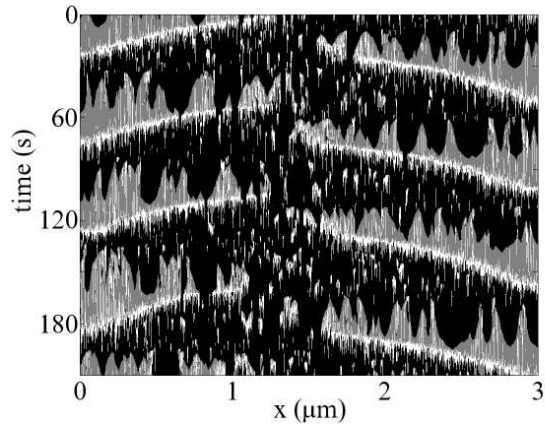} 
 \caption{Space-time plot for occupancy of a single membrane filament
 with $n_{\rm max}=12$ and $N_{\rm c}=2$. Black areas are empty, gray shows 
MinD and white is the MinDE complex.}  \label{p1}
\end{figure}

When a polymer has a chance to form in a region with little MinE, fast cooperative
binding means the polymer grows rapidly in both directions, towards the centre
and the pole of the cell. MinD polymers in regions with high MinE concentrations
do not grow to a significant length, as the MinE prevents further cooperative
binding and causes dissociation from the membrane. From figures \ref{standard}
and \ref{p1} we can see that near mid-cell there are a large number of small
patches of MinD, which are short in length and short-lived. These are quickly
occupied by MinE and displaced from the membrane. Figure \ref{p1} shows that
the pattern of each individual filament follows that of both filaments taken
together.

As MinE relocates from the other end of the cell by cytoplasmic diffusion,
it will tend to bind to the membrane at the first encountered region of elevated
MinD concentration. Hence, as can clearly be seen in figure \ref{standard}($b$),
a tightly localized region of high MinE concentration (the ``MinE ring'')
typically accumulates at the end of the region of high MinD concentration.
Since MinD forms polar zones, the MinE ring is nucleated close to mid-cell
and thereafter moves towards the pole, via detachment, diffusion and reattachment,
as the MinD region shrinks. Although the different filaments are independent,
they are disassembled simultaneously since MinE binds equally to each.

\textit{Time-averaged concentrations:}
\begin{figure} 
 \includegraphics[width=0.5\textwidth]{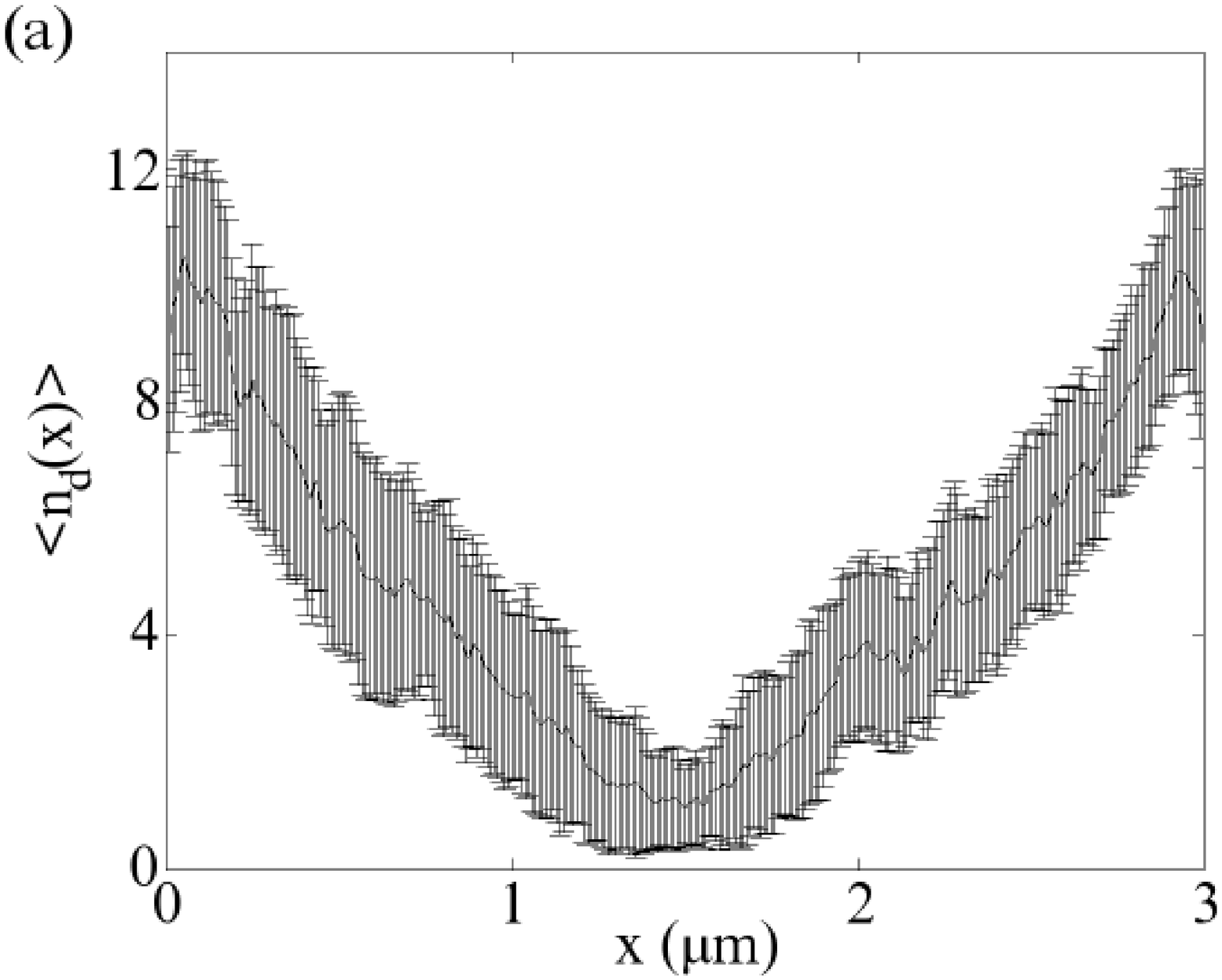} 
 \includegraphics[width=0.5\textwidth]{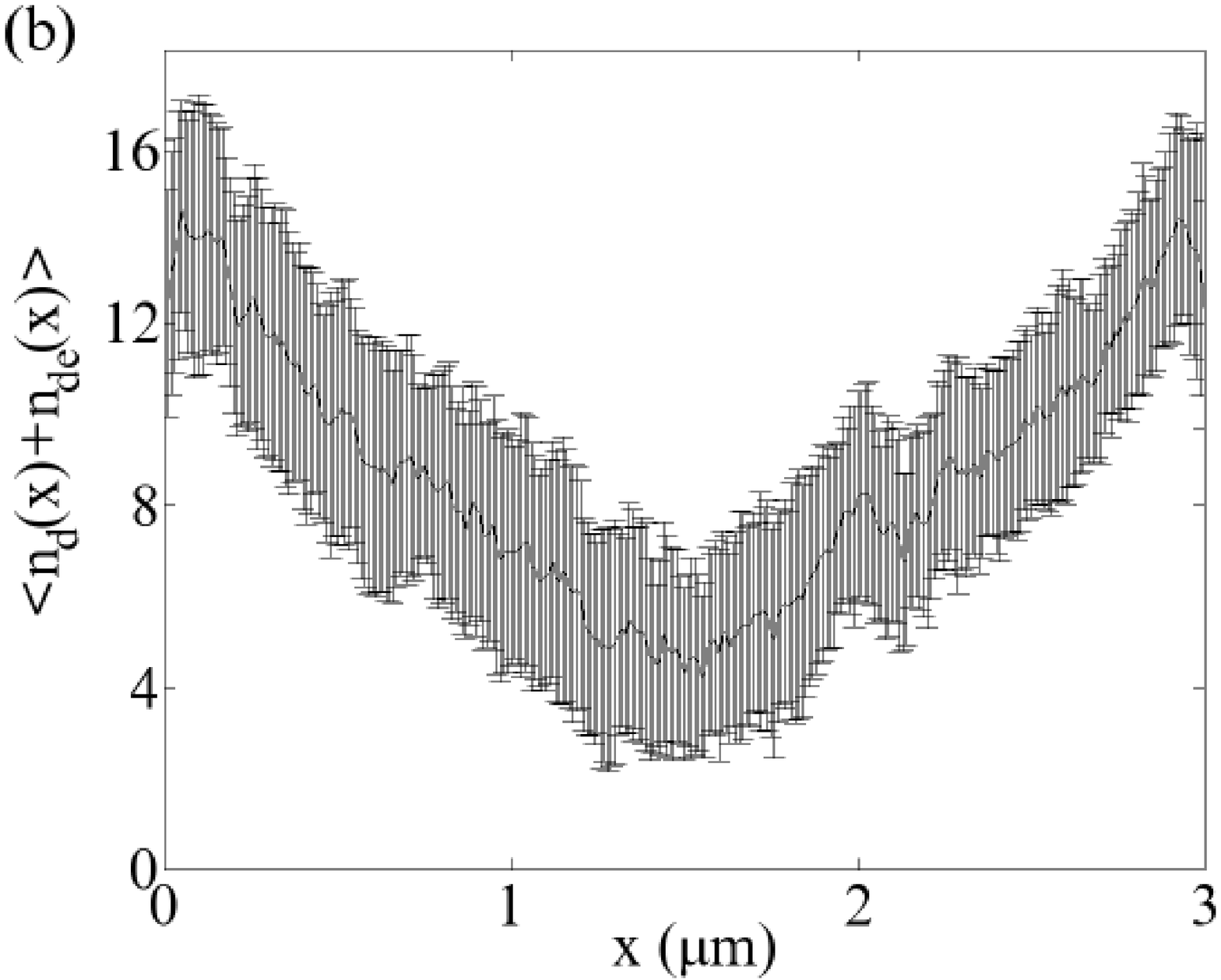} 
 \includegraphics[width=0.5\textwidth]{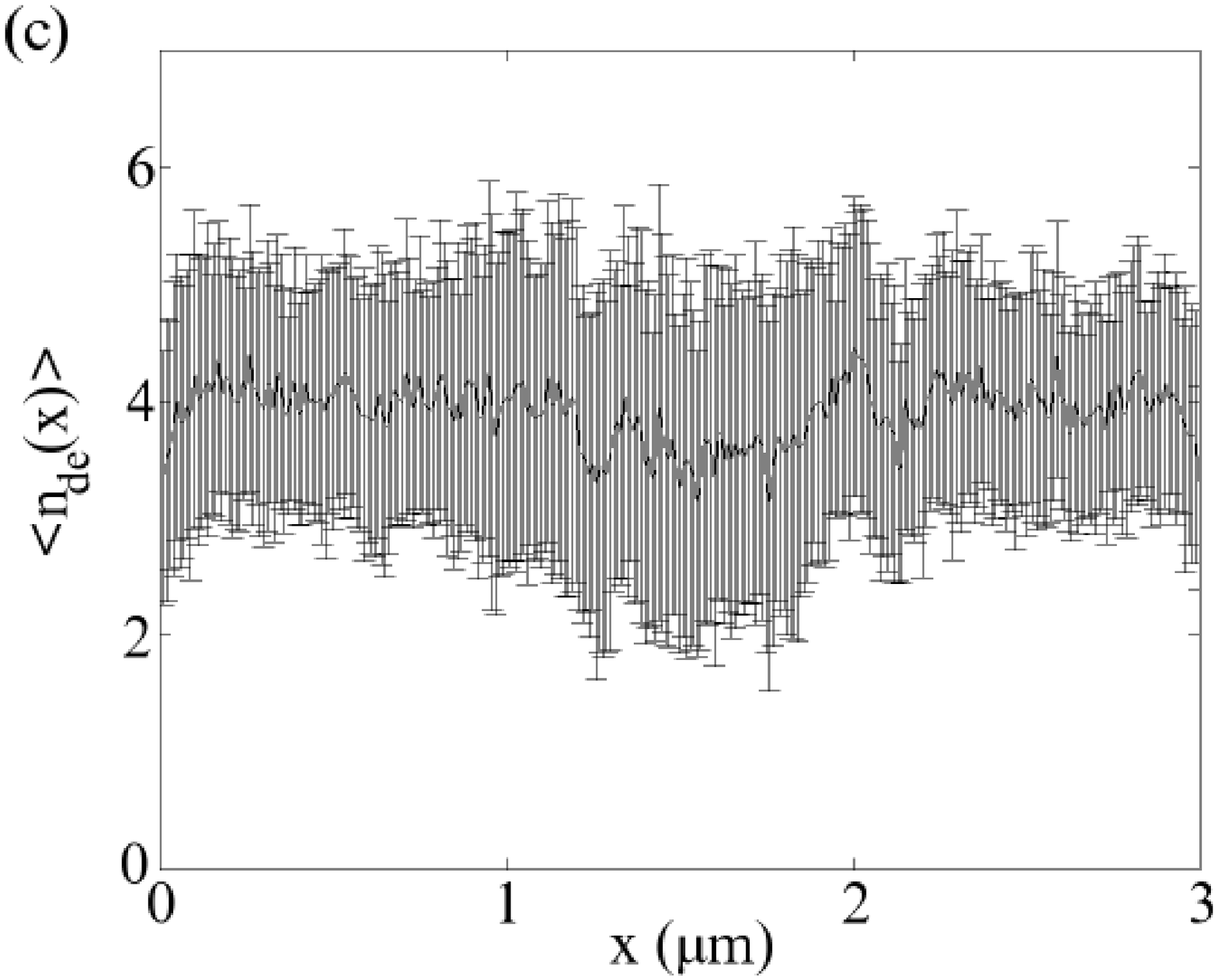} 
 \caption{The time-averaged amount and oscillation-to-oscillation 
variability of ($a$) MinD not including the MinDE complex, ($b$) 
total MinD and ($c$) MinDE present on the membrane as a function of 
position along the cell. See text for details of the averaging procedure.}
 \label{<>}
\end{figure}
Oscillation cycles were identified as periods between the MinE ring reaching
one cell pole. This was done manually by looking at $n_{\rm de}^1$, identifying
times where the occupancy was high for an extended period, and defining the
end of the cycle as the time when the occupancy dropped to below $N_{\rm
c}n_{\rm max}/2$. For each of the Min proteins, the membrane density as a
function of position was averaged over each oscillation cycle. Figure \ref{<>}
shows the mean and standard deviation of these profiles over a large number
of oscillation cycles. We can see that fluctuations in our stochastic model
do not destroy the biologically important midcell concentration minima for
MinC and MinD.

The key result for cell division is that the concentration of MinC (which
in our model is quantified by $n^i_{\rm d}$) is maximized at the ends of
the cell, suppressing Z-ring formation at these locations. The total amount
of membrane-bound MinD, including the MinDE complex ($n^i_{\rm d}+n^i_{\rm
de}$), also has a minimum around the cell centre and maxima at the cell ends.
This result is in good agreement with experimental observations \cite{Kruse2}.
In our model, the average amount of membrane-bound MinE is roughly constant
along the length of the cell, although with large fluctuations. This contrasts
with other models which have a minimum \cite{Kruse1,Huang,Kruse2,Pavin} or
maximum \cite{Howard1} for membrane MinE at the cell centre. This profile
has not been measured experimentally. Such a measurement could potentially
distinguish between the various models.

\textit{Variation of period with protein numbers:}
\begin{figure} 
 \includegraphics[width=0.5\textwidth]{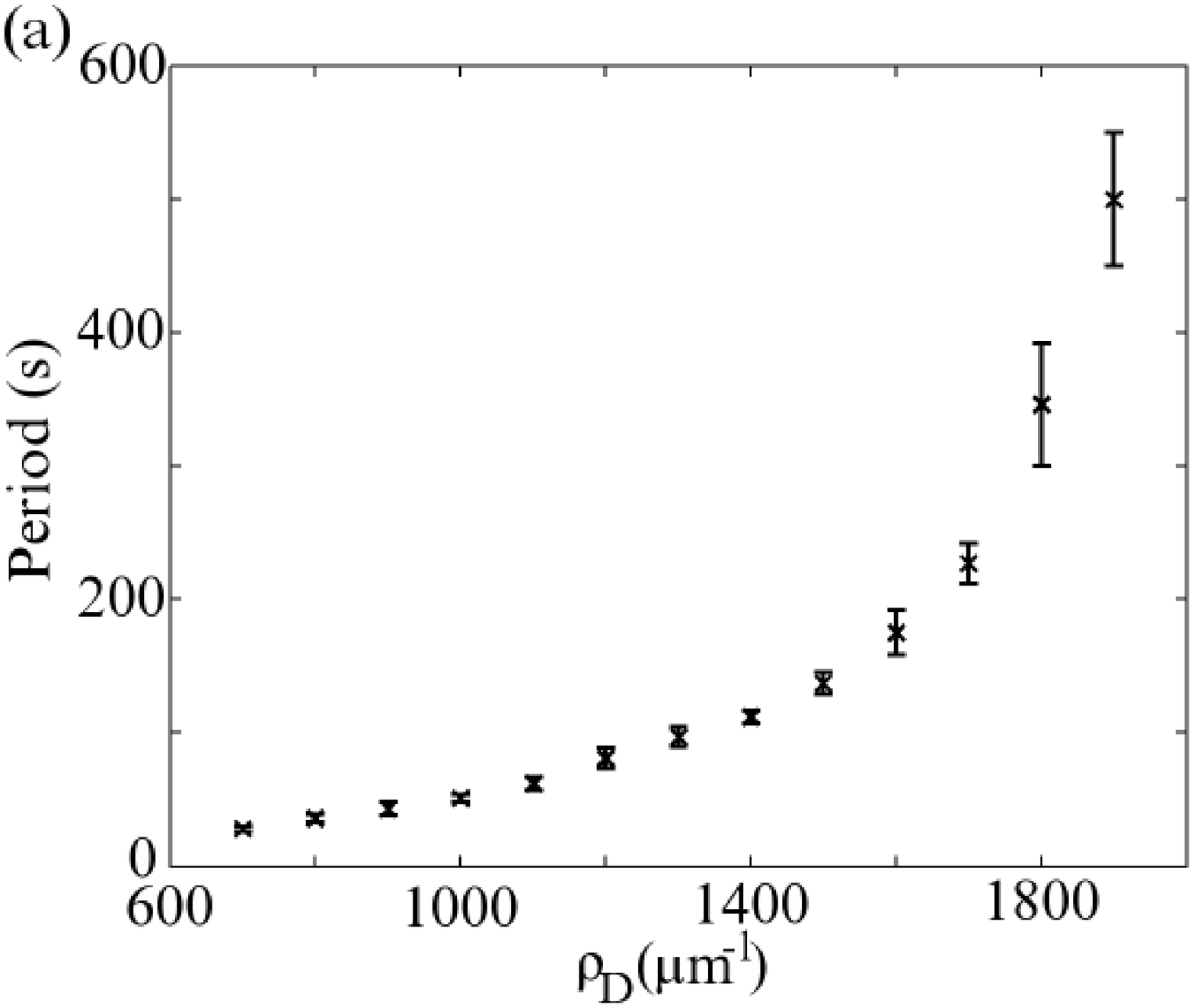} 
 \includegraphics[width=0.5\textwidth]{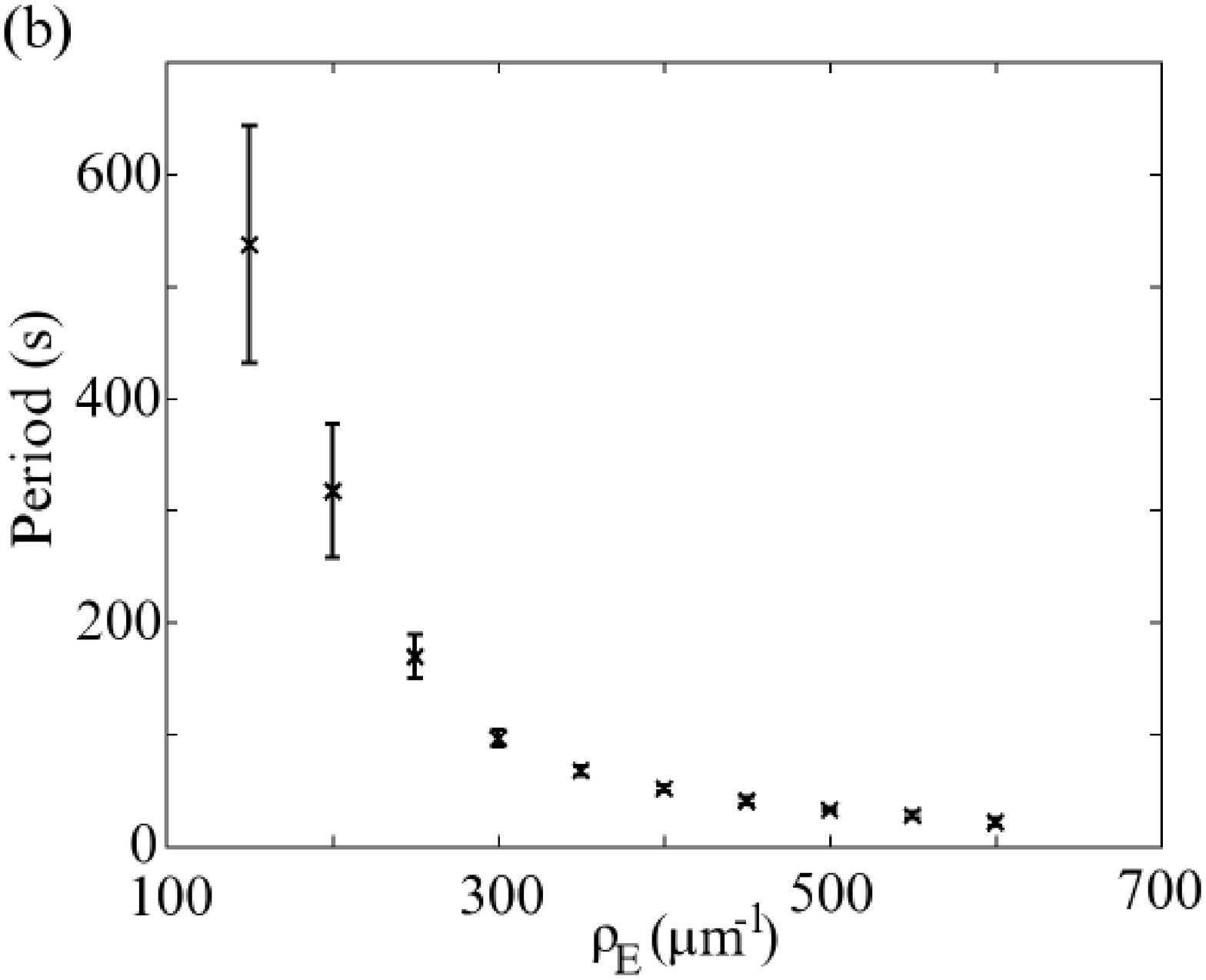} 
 \includegraphics[width=0.5\textwidth]{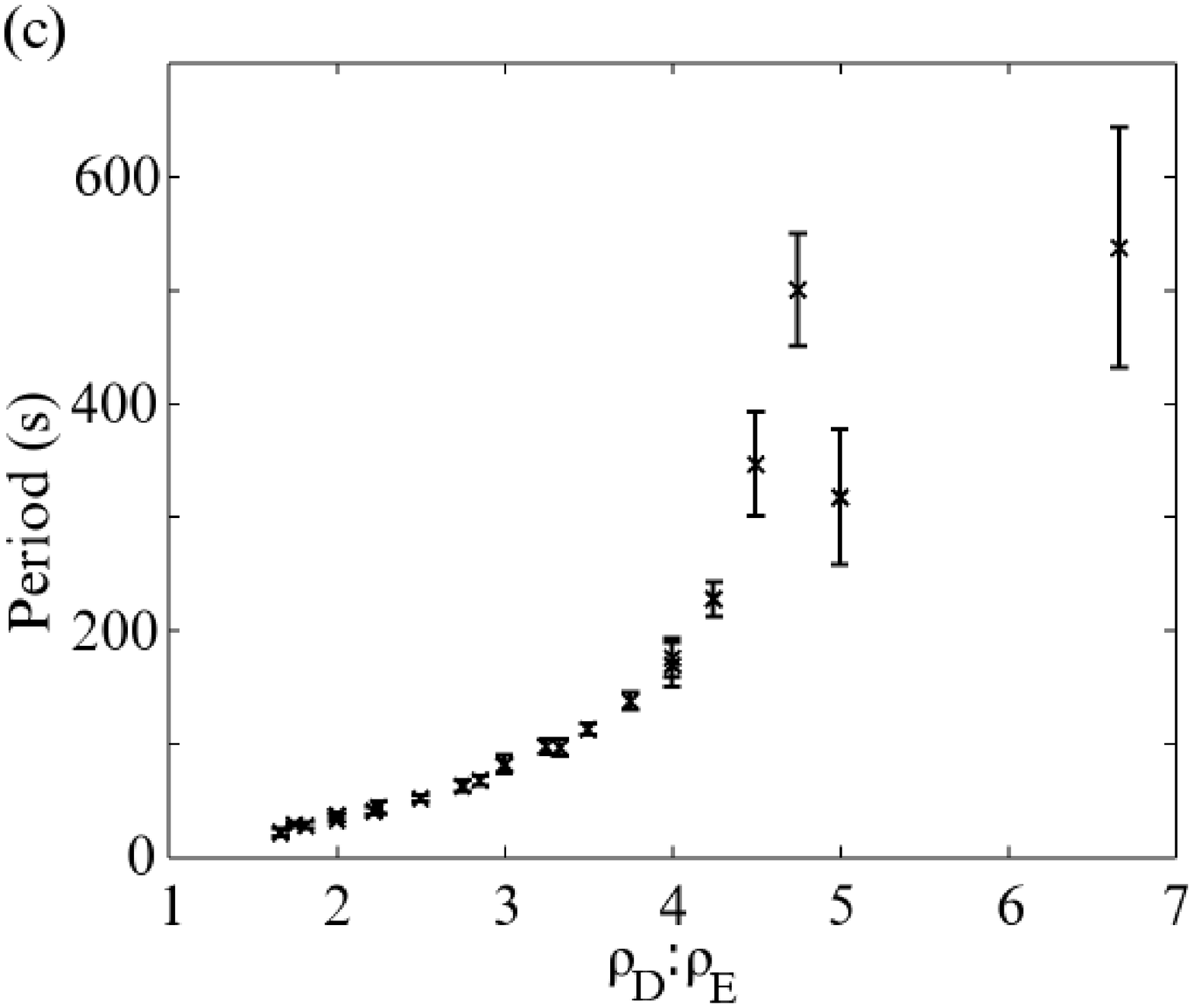}
 \includegraphics[width=0.5\textwidth]{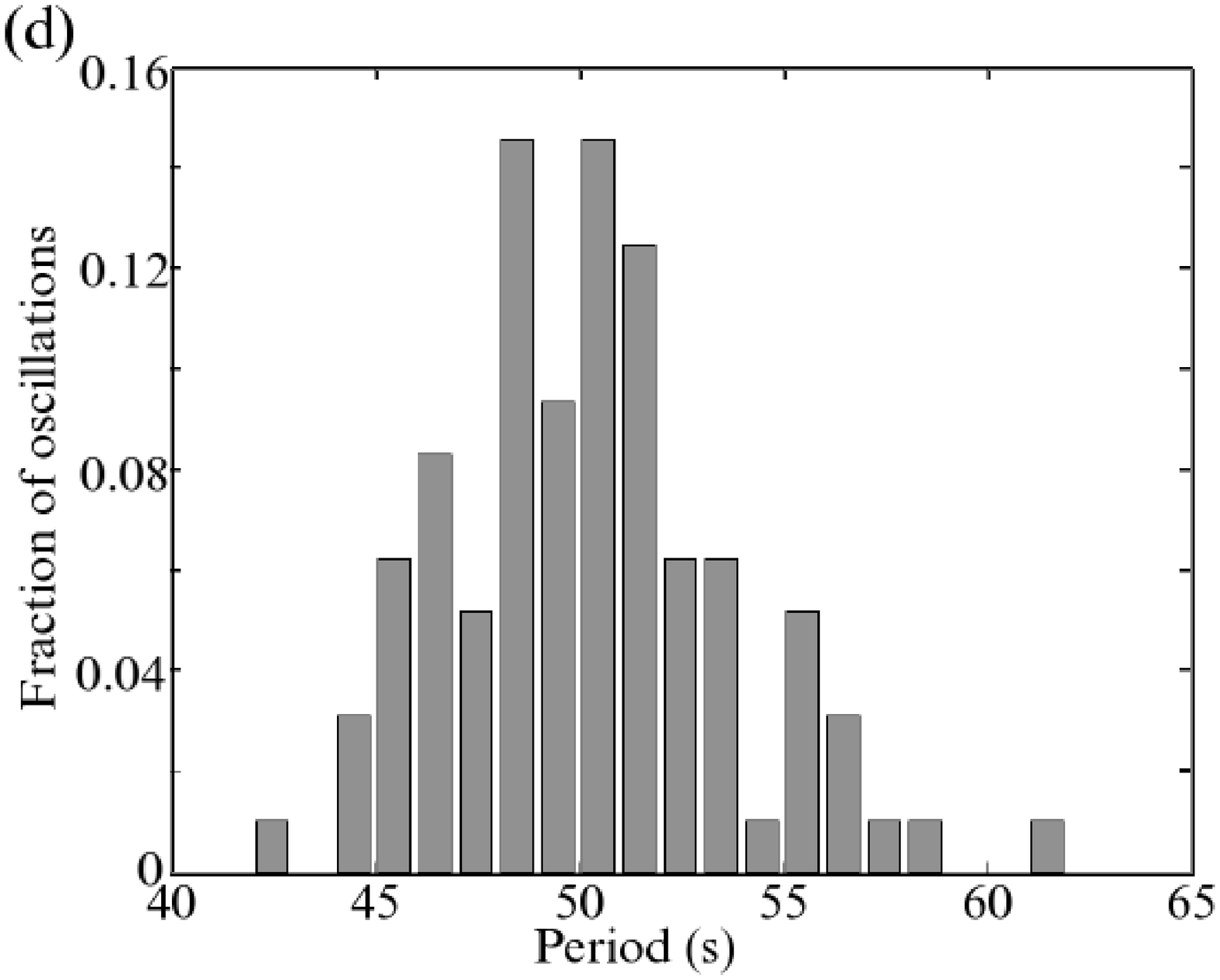} 
 \caption{Variation of oscillation period when varying ($a$) $\rho_{\rm D}$
with $\rho_{\rm E}=400\mu m^{-1}$, ($b$) $\rho_{\rm E}$ with $\rho_{\rm
D}=1000\mu m^{-1}$, and ($c$) $\rho_{\rm D}:\rho_{\rm E}$ ratio. ($d$) Distribution
of 100 periods for the case with $\rho_{\rm D}=1000\mu m^{-1}$ and $\rho_{\rm
E}=400\mu m^{-1}$. The distribution is similar in other cases. In those
cases where the observed period is less than about $100s$, the standard
deviation is always close to $10\%$ of the period.}  \label{ratio} 
\end{figure}
Figure \ref{ratio} shows that the oscillation period increases with
increasing MinD concentration, and decreases with increasing MinE
concentration. This is consistent with experimental observations
\cite{RdB}. The range of periods supported in this model also covers 
that observed \textit{in vivo}, where the variation is likely due to the
fluctuations in protein copy numbers between different cells.

Oscillations occur for a fairly large range of $\rho_{\rm D}:\rho_{\rm E}$
ratios, but cut off when the $\rho_{\rm D}:\rho_{\rm E}$ ratio drops below
about $1.6$. At these concentration levels, MinD filaments are unable to
grow to a significant length because they are removed from the membrane too
quickly. At the opposite end of this scale there is no sharp transition;
increasing $\rho_{\rm D}:\rho_{\rm E}$ causes the polar zones to extend further
into the opposite half of the cell. Above the range shown in figure \ref{ratio}($a$),
the ``polar zone" effectively extends for the whole length of the cell and
MinE is unable to empty the membrane. \label{res}

\textit{Filamentous cells:}
Observations of filamentous cells which are unable to divide have revealed
regularly spaced bands of MinD with accompanying MinE rings \cite{RdB,Fu,Hale}.
This is strong evidence in favour of a dynamic instability mechanism for
the oscillations, since the presence of bands supports the existence of a
characteristic wavelength for the dynamics independent of the cell length.
\begin{figure}
 \includegraphics[width=0.5\textwidth]{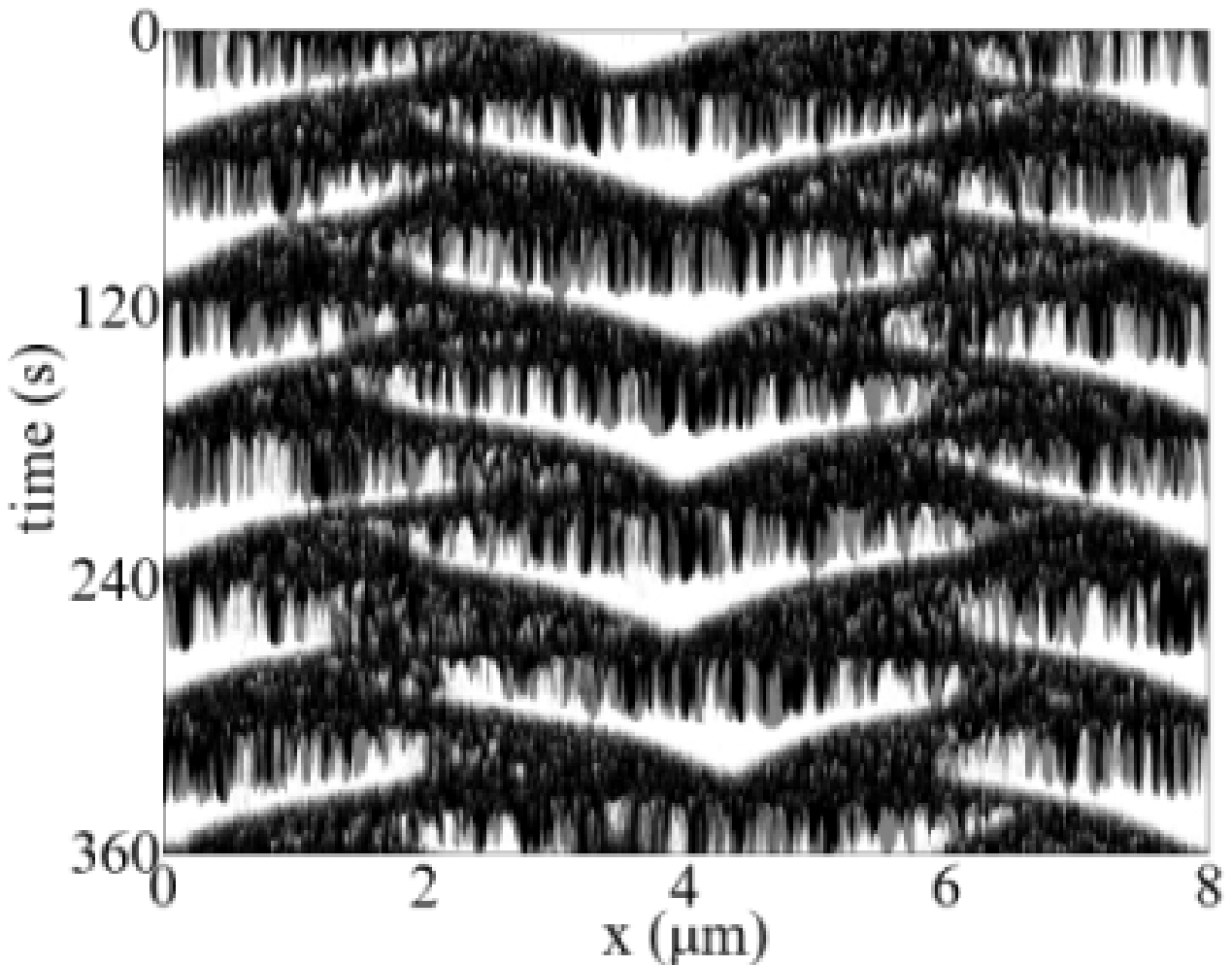} 
 \includegraphics[width=0.5\textwidth]{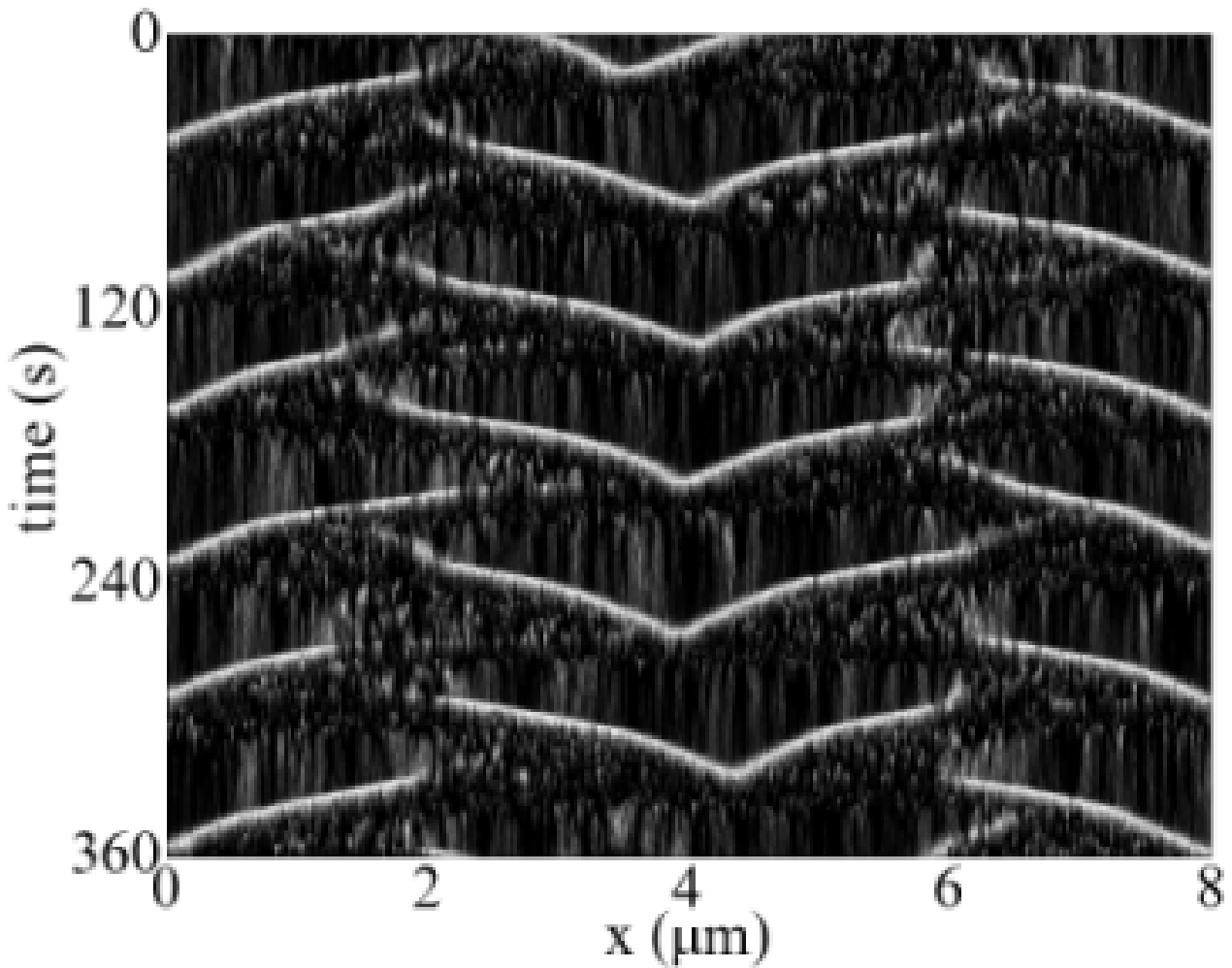} 
 \includegraphics[width=0.5\textwidth]{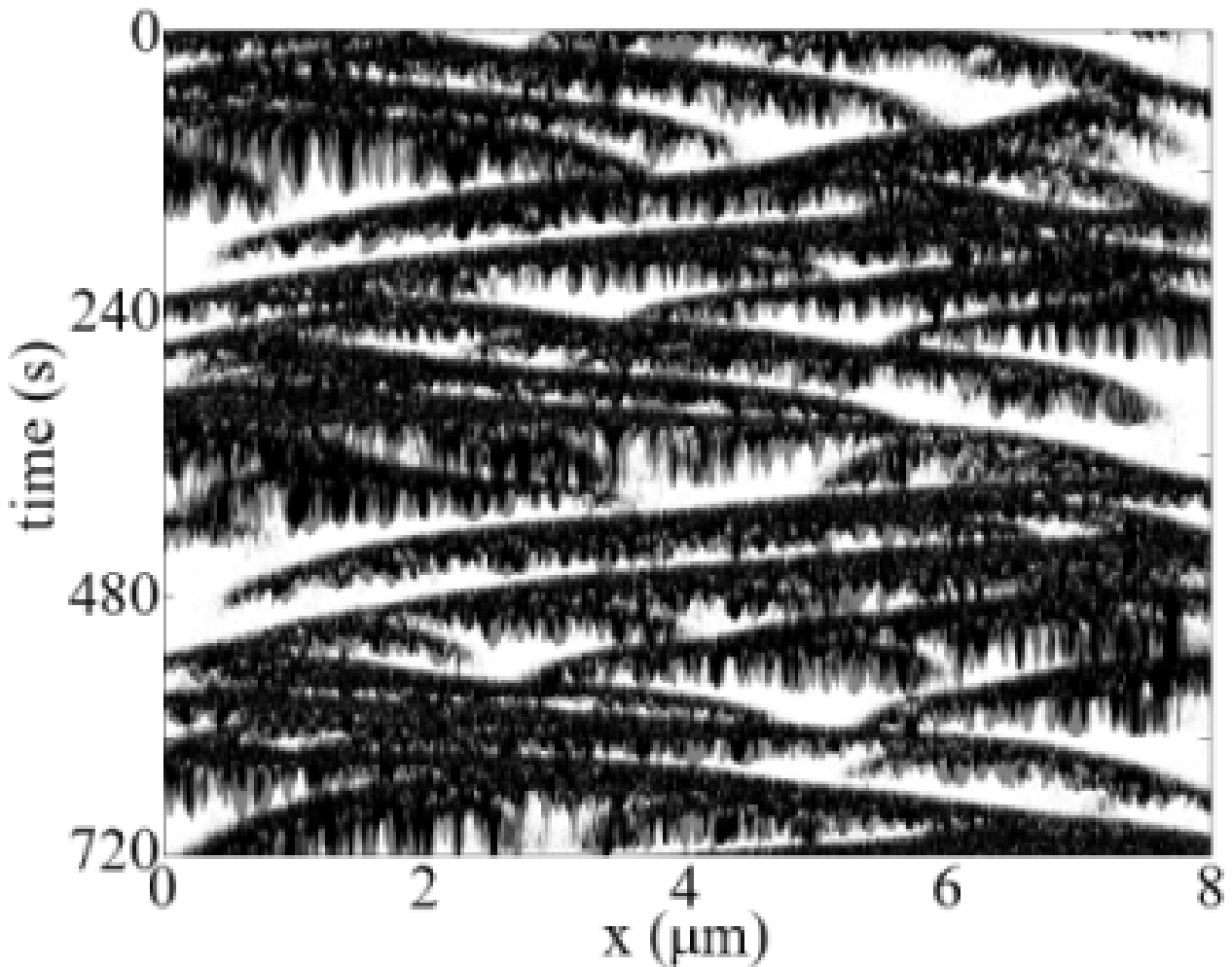} 
 \includegraphics[width=0.5\textwidth]{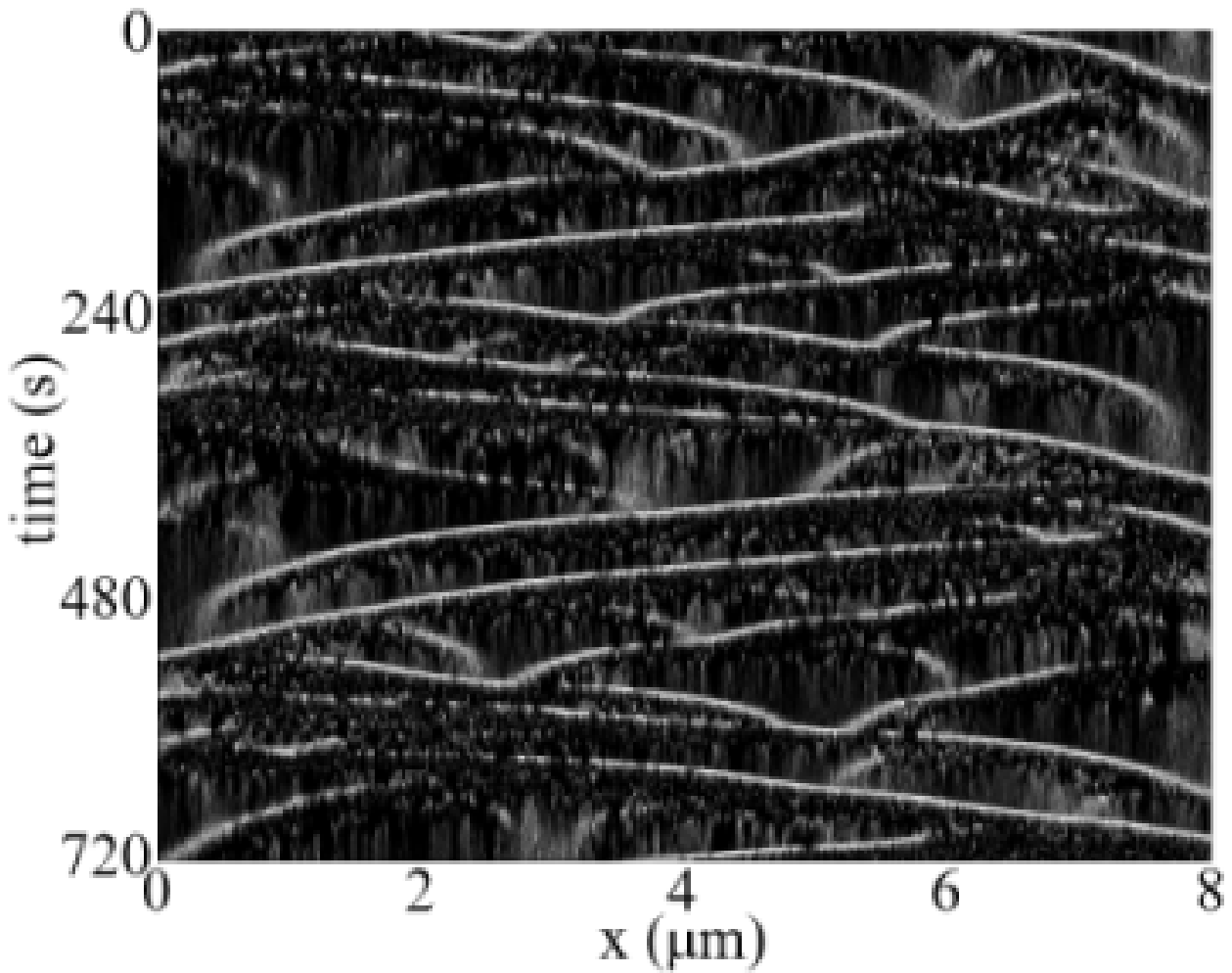} 
 \caption{Membrane occupancy in $8\mu m$ cells, showing both periodic and
 more disordered dynamics. Plots on the left show membrane MinD. Plots on
 the right show the corresponding MinDE complex distribution.} 
 \label{filament} 
\end{figure}
Figure \ref{filament} shows the results of simulations of our model
performed in longer cells. In some cases, periodic oscillations
with a number of MinD bands are observed, with the number of bands 
increasing with the cell length. In other cases, several 
regularly spaced bands form, but these all advance towards the same
cell pole. In these cases the dynamics is more disordered. Such
disordered behaviour has not yet been reported experimentally. However
our model predicts that, while periodic behaviour may be seen over
some intervals of up to 30 minutes, many filamentous cells will also
have periods of irregular dynamics or switch between single and double
banded oscillations. Such irregularity is perhaps not surprising given
the stochastic nature of our model, and would certainly be interesting
to search for experimentally.

\textit{Variation of period with length:}
\begin{figure} 
 \centering 
 \includegraphics[width=0.5\textwidth]{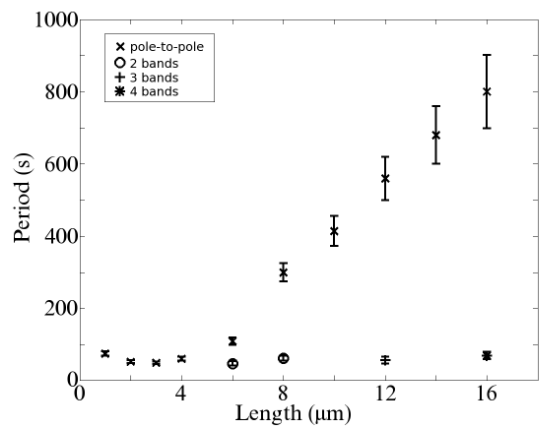} 
 \caption{Variation of period with cell length, $L$.} 
 \label{L} 
\end{figure}
Figure \ref{L} shows the variation of oscillation period with cell length,
while keeping the protein concentrations constant so the total protein number
increases proportional to $L$. Over the range $1\mu m\leq L< 6\mu m$, where
only single banded pole-to-pole oscillations are observed, the period remains
approximately constant as the length is varied. Experimental evidence \cite{Kruse2}
is that any change in the period with length \textit{in vivo} is much smaller
than the variation in period at constant length, which is presumably due
to concentration levels differing between individual cells. When multiple
oscillation bands are observed in longer cells, beginning at about $L=6\mu
m$, their period is similar to that of the pole-to-pole oscillations in shorter
cells. 

In the case of disordered behaviour it is more difficult to identify a characteristic
period in the observed dynamics. However, the dynamics is often dominated
by the bulk of the MinD sweeping regularly from one pole to the other, and
we use this to find the dominant period of oscillation. For example in the
lower panels of figure \ref{filament}, $t=230s$ to $t=530s$ would be considered
to be one period. The period of this type of oscillation increases linearly
with cell length, in contrast to the roughly constant period observed for
$L<6\mu m$.

We believe this difference in behaviour is the result of two qualitatively
different types of dynamics. In short cells continuous MinD zones form between
the cell pole and the MinE ring at midcell. These polar zones have a characteristic
time associated with their disassembly regardless of cell length, which gives
rise to the constant period of these oscillations. In long cells, the pole-to-pole
oscillations are made up of MinD bands at intervals of $3-4\mu m$ that migrate
across the cell through disassembly on one side and growth on the other.
These bands always move at a particular speed, giving the linear increase
in period with cell length.

However, the linear relationship intersects the $L$-axis at approximately
$L=4\mu m$. MinD bands in long cells travel less than the full length of
the cell, because they form slightly away from the previously occupied pole
and because once these MinD bands approach the other cell pole, polar zones
similar to those in short cells form. So we can consider the oscillation
period in long cells to be made up of two parts: the time to disassemble
the polar zones, which is the oscillation period in short cells, plus the
time taken for the MinD bands to travel twice across about ($L-4$)$\mu m$
of the cell.


\section{Oscillations during cell division}

Now that we have established that our model reproduces the \textit{in
vivo} behaviour of the Min system, we use the model to investigate the
Min dynamics during cell division. We investigate two mechanisms to
simulate the closing septum, and examine how the Min oscillations are
altered both during this process and once the daughter cells have
separated. In particular we would like to study the distribution of
the numbers of the Min proteins in each daughter cell, as this has not
yet been measured experimentally.

\textit{Model A:} Let $t$ be the time since invagination began and $T$
be the total time from when invagination begins to when there is no
longer a cytoplasmic connection between the daughter cells. Over a
length, $2l$, centred at $x=L/2$, we assume that the invagination of
the cell membrane causes ``compression" of the cytoplasm, making
diffusion more difficult. As a result of this compression, diffusion
decreases to zero in this region by time $T$, and unless otherwise 
stated we assume that this decrease occurs quadratically with time. 
In model A, we therefore employ a reduced diffusion probability, 
$D'(t)dt/{(dx)}^2$, in the region $L/2-l\leq x \leq L/2+l$ with
\begin{equation}
D'(t)=D_{\rm 0}\left(\frac{T-t}{T}\right)^2,
\end{equation}
and where $D_{\rm 0}$ is the cytoplasmic diffusion constant in the rest of
the cell.

Model A provides a simple way to implement the division process. However
it is perhaps unrealistic to assume that diffusion is reduced equally over
the whole range $2l$, particularly as there is little clear evidence for
this ``compression" of the cytoplasm. This model also neglects the importance
of the direction of diffusive motion, whether towards or away from the septum
and into a narrower or wider region. We therefore also investigate a second,
possibly more realistic, model.

\begin{figure} 
 \centering 
 \includegraphics[width=0.6\textwidth]{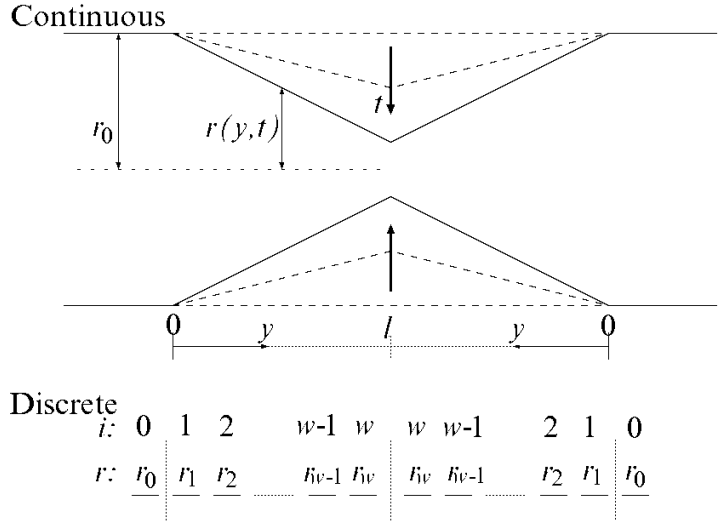} 
 \caption{Schematic of the Model B septal region.} 
 \label{sep} 
\end{figure}

\textit{Model B:} Figure \ref{sep} shows a schematic of this
mechanism. Let $y$ be the distance from the outer edge of the
narrowing region measured towards the centre.  We assume that the cell
radius decreases linearly with $y$, and that the radius closes
linearly with time:
\begin{equation}
r(y,t)=r_{\rm 0}\left(1-\frac{y}{l}\frac{t}{T}\right).
\label{rxt}
\end{equation}

Equation (\ref{rxt}) discretizes to give
\begin{equation}
r_i(t)=r_{\rm 0}\left(1-\frac{i-1}{w}\frac{t}{T}\right), \ i\geq 1.
\end{equation}
where $w$ is the number of sites in the contracting region and $i$ is
the site number counting from the polar end of this region. The presence
of the $-1$ in the numerator simply reflects a choice in the discrete model
of precisely where the invagination begins in space. The
probability of diffusing into the next site towards the cell centre
is assumed to vary with the ratio of the cross-sectional areas $A_i$, 
where $A_i\propto r_i^2$, since the narrowing cell radius may restrict the
mobility of protein particles close to the membrane.  This
is equivalent to reducing the diffusion probability towards the
septum from site $i$ to site $i+1$, $D_i(t)dt/{(dx)}^2$, according to
\begin{eqnarray}
D_i(t)&=&D_{\rm 0}\frac{A_{i+1}(t)}{A_{i}(t)} \nonumber \\ &=&
D_{\rm 0}\frac{\left(1-\frac{i}{w}\frac{t}{T}\right)^2}
{\left(1-\frac{i-1}{w}\frac{t}{T}\right)^2},
\ i=1, \ldots ,w.
\end{eqnarray}
The probability of diffusion away from the septum is unchanged at
$D_{\rm 0}dt/{(dx)}^2$.

Unless otherwise stated we use $T=300s$ and $l=0.1\mu m$ (estimated from
\cite{lutk_ftsz}) or $w=10$.

\subsection{Results}

\begin{figure} 
 \includegraphics[width=0.5\textwidth]{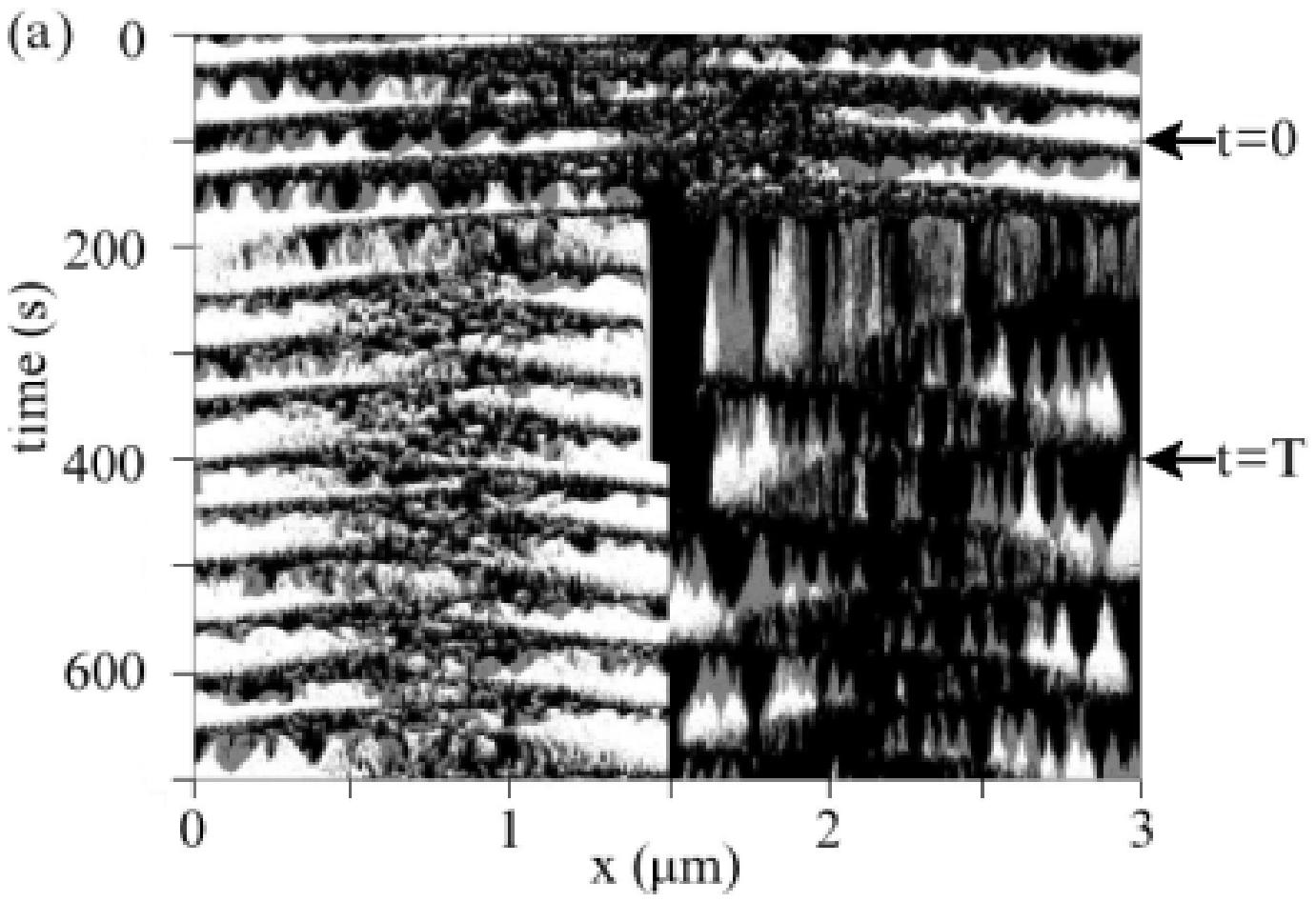} 
 \includegraphics[width=0.5\textwidth]{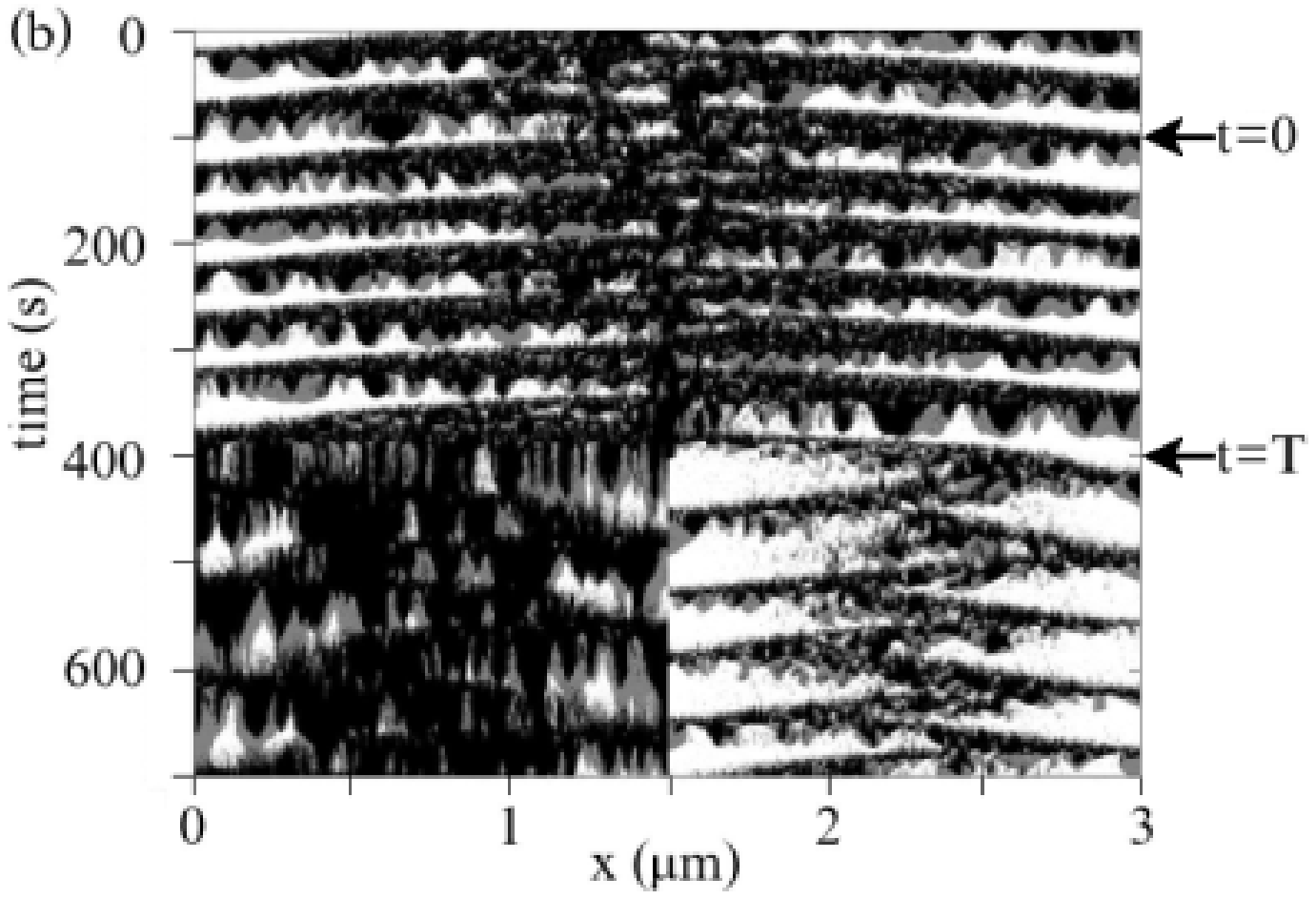} 
 \caption{Space-time plot showing MinD oscillations in a dividing cell, for
 ($a$) model A and ($b$) model B. The division process begins at the point
 marked $t=0$ and ends at $t=T$. The grey-scale used is the same as in figure
 \ref{standard}.}
 \label{div} 
\end{figure}
Oscillations are initially unaffected as diffusion through the septum 
is reduced. Then at some later time diffusion through the septum cuts off
sharply. After this time the two daughter cells are effectively independent,
even though there remains a connection through the cytoplasm. This cut-off
time varies between models but is approximately independent of the density
distributions at $t=0$. In model A, pole-to-pole oscillations cease relatively
quickly, after approximately one minute. In model B, where the diffusion
rate is on average greater because of the additional spatial variation, oscillations
continue with little obvious alteration for about 270 seconds. 

At the centre of the cell there is a region where the membrane remains empty,
which appears at about the time when pole-to-pole oscillations are disrupted.
Possibly the reduced diffusion probability makes it less likely that any
proteins will be able to enter these sites, and thus reoccupy the membrane.
For model A this includes about half of the contracting region, as can be
clearly seen in figure \ref{div}($a$). At $t=T$ the empty central region
is quickly reoccupied because we restore the diffusion rate to $D_{\rm 0}$
(except at $x=L/2$) and proteins can once again access these sites. For model
B, the empty region extends only over a few sites at the centre of the cell
and appears much later during division. Again this is due to the greater
diffusion rates in model B.

\begin{figure} 
 \includegraphics[width=0.5\textwidth]{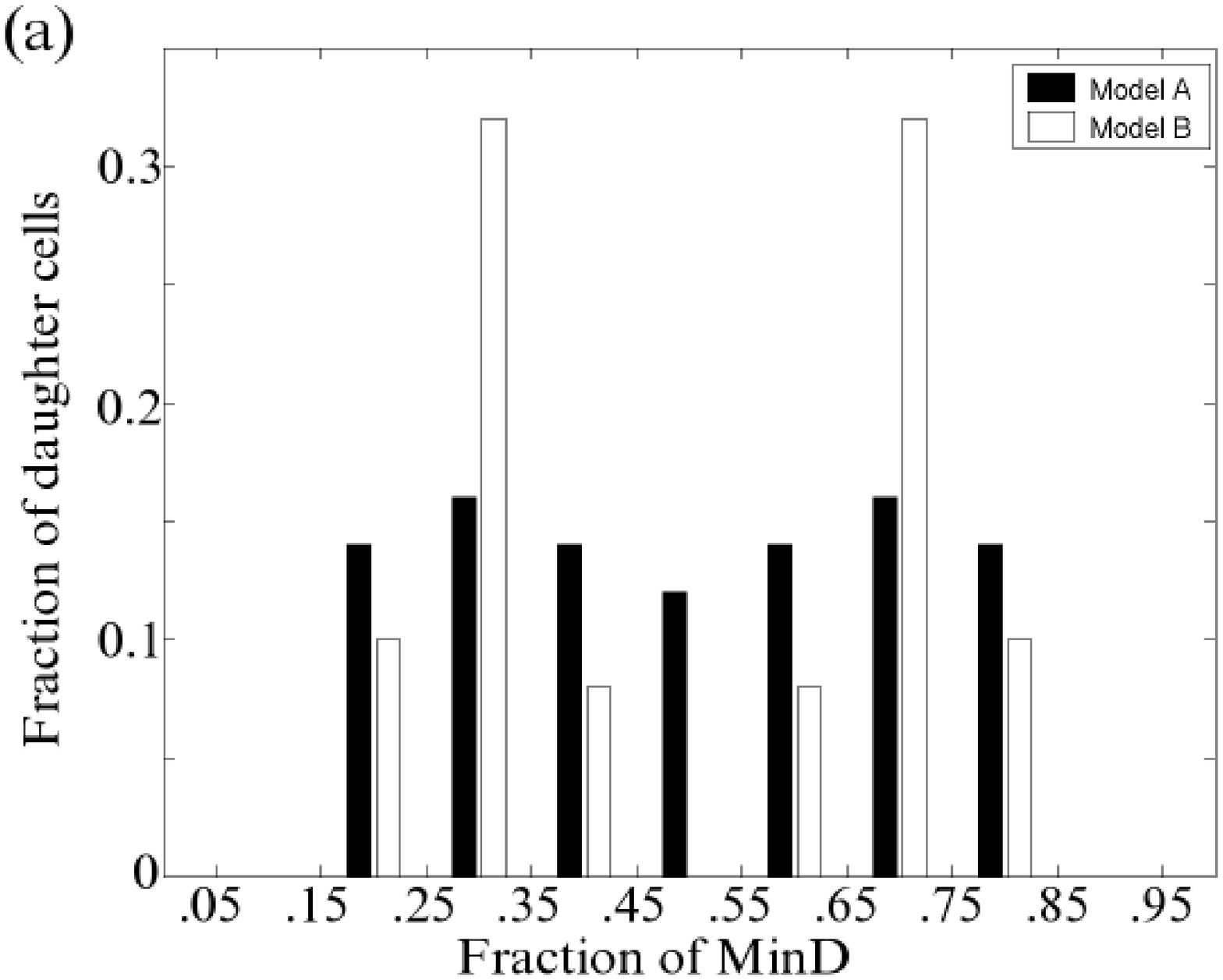} 
 \includegraphics[width=0.5\textwidth]{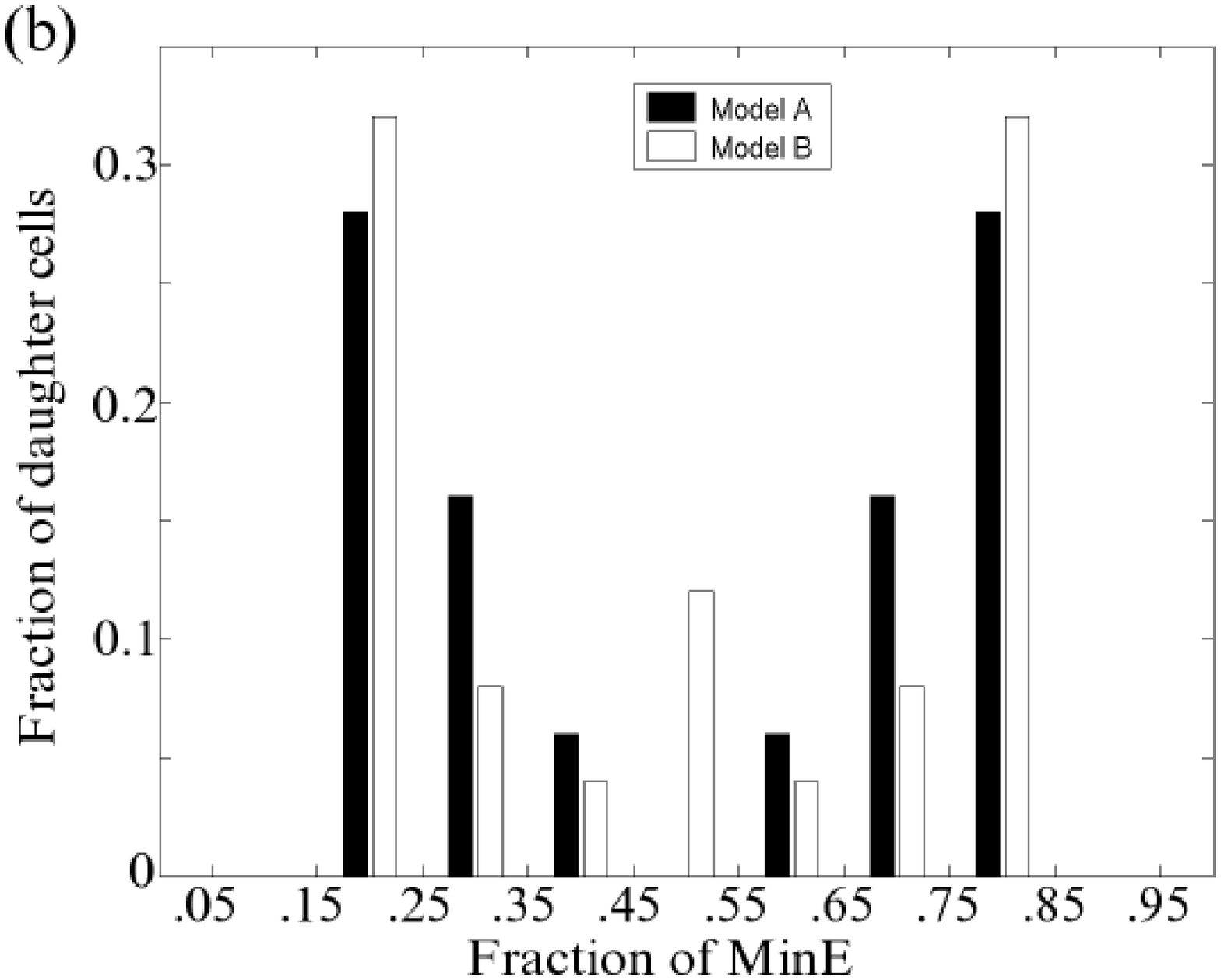}
 \includegraphics[width=0.5\textwidth]{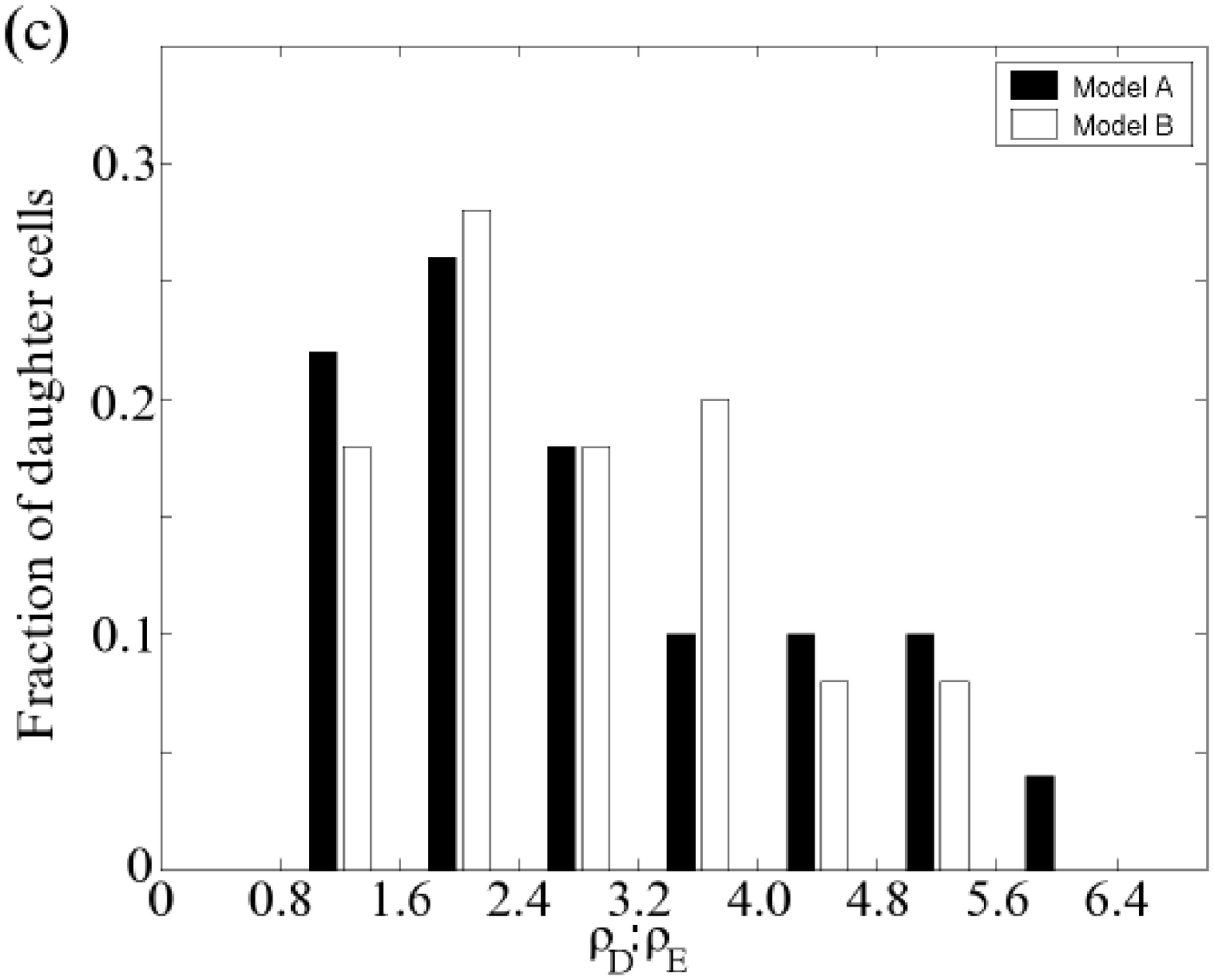}
 \caption{Comparison of the division models A and B, showing the distributions
 of the fraction of ($a$) MinD molecules and ($b$) MinE molecules from the
 parent cell, and ($c$) of $\rho_{\rm D}:\rho_{\rm E}$ ratios, in the daughter
 cells.}
 \label{AB} 
\end{figure}

Protein numbers in the daughter cells vary from 85\% to 15\% of the total
in the parent cell for both MinD and MinE. This range is the same as the
variation in protein numbers in each half of the parent cell during normal
pole-to-pole oscillations. Figure \ref{AB} compares the daughter cell distributions
between the two models. In both cases, the MinE distribution peaks at high
and low concentrations. In model A, an equal distribution into the two daughter
cells is never observed. The $\rho_{\rm D}:\rho_{\rm E}$ ratios in daughter
cells are also similar in the two models. Only the MinD distribution shows
a significant difference between the two models. In model A, all concentrations
are approximately equally likely. In model B, however, copy numbers in the
daughter cells between 25-35\% and 65-75\% of the total from the parent cell
are strongly favoured and a 50\%-50\% split is never observed.

The $\rho_{\rm D}:\rho_{\rm E}$ ratio in daughter cells ranges from about
1.3 to 6. Those daughter cells with $\rho_{\rm D}:\rho_{\rm E}<1.6$, approximately
20\% of the total produced in our simulations, cannot support pole-to-pole
oscillations because MinD is unable to form sufficiently long filaments on
the membrane. This is consistent with our results in section \ref{res}. All
daughter cells with $\rho_{\rm D}:\rho_{\rm E}>1.6$ did have Min oscillations.
However when the protein copy number is low, polar zones are less dense and
fluctuations become more significant in the dynamics. 

\begin{figure} 
 \centering 
 \includegraphics[width=0.6\textwidth]{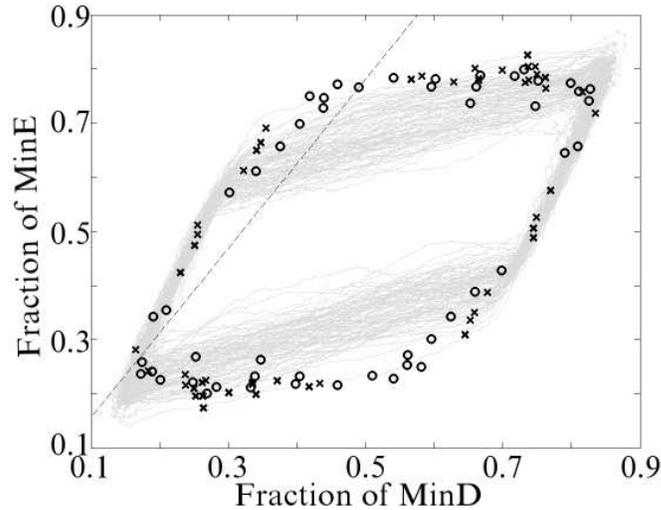} 
 \caption{Points show the fraction of MinE from the parent cell in each daughter
 cell plotted against the fraction of MinD in the same cell. After division,
 these fractions are of course constant for each daughter cell. Circles represent
 model A, and crosses model B. The gray lines show the fraction of each protein
 in one half of the parent cell as a function of time during pole-to-pole
 oscillations. This is slightly disordered due to fluctuations. The dashed
 line indicates $\rho_{\rm D}:\rho_{\rm E}=1.6$. Daughter cells to the left
 of this line do not have pole-to-pole oscillations.}
 \label{loop} 
\end{figure}

If we plot the fractions of MinE and MinD in the same half of the parent
cell as a function of time as pole-to-pole oscillations take place, the result
is a cycle as shown in figure \ref{loop} (gray lines). During the division
process, the Min protein dynamics are of course altered. Hence, as can be
seen in Figure \ref{loop}, the data points showing the fraction of the proteins
ending up in the daughter cells lie on another closed loop which is similar,
though not identical to, the cycle of the parent cell. We can also see that
both models A and B produce daughter cells with protein fractions that lie
on the same closed loop. 

\subsection{Robustness}

\begin{figure} 
 \centering 
 \includegraphics[width=0.6\textwidth]{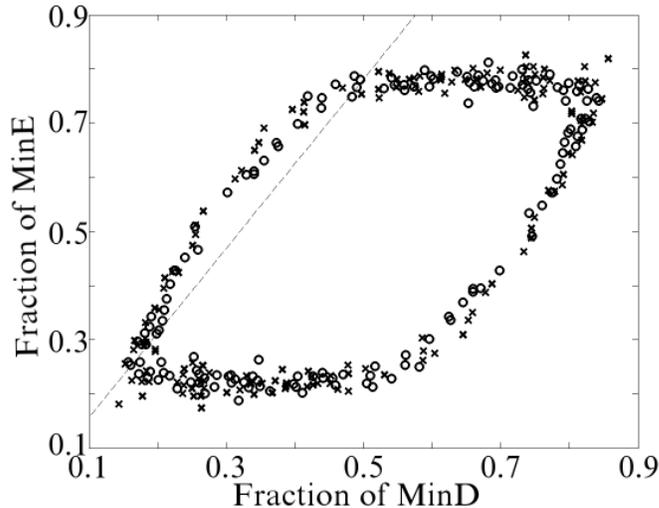} 
 \caption{As for figure \ref{loop}, but with data added for different $w$
 and $T$ values and different functional time dependences.}
 \label{loop2} 
\end{figure}

The results presented above appear to be general and are qualitatively the
same under a number of changes (discussed below) to the division models.
No systematic trends were observed when varying any of the parameters in
either of the models. In fact when additional data from these perturbed models
is added to the data from figure \ref{loop}, all the data points continue
to lie on the same loop (see figure \ref{loop2}).

\textit{Width of contracting region:} Increasing $w$ means that the pole-to-pole
oscillations of the parent cell are disrupted sooner, because the cumulative
probability of diffusion from one half of the cell to the other is reduced.
Conversely, if $w$ is reduced oscillations in the parent cell will continue
later into the division process. However there is no obvious effect on the
protein numbers in the daughter cells when $w$ is increased to $20$ or reduced
to $5$.

We have also tested the case where diffusion is reduced only when crossing
from one half of the cell to the other, a limiting case of our earlier models.
The observed distribution of protein numbers into the daughter cells is again
the same.

\textit{Form of time-dependence:} We have tested model B with $r(y,t)$ decreasing
quadratically with $t$, and model A with linear time dependence. Again the
behaviour is qualitatively the same. The time at which oscillations cease
is earlier if the diffusion probability decreases more rapidly with $t$,
and later if the diffusion probability decreases more slowly. However the
distributions of the Min proteins into the daughter cells are unaffected.

\textit{Division time, $T$:} Again the distribution of Min proteins into
the daughter cells showed no systematic changes. The time at which the oscillations
in the parent cell ceased appeared to vary linearly with $T$, so it always
occurred at a particular value for the diffusion probability $D'$ or $D_w$.
$T=150s$ and $T=450s$ were tested in addition to $T=300s$.

\textit{Stochastic vs. continuous models:} We also implemented a similar
mechanism to model A into continuous partial differential equation models
adapted from \cite{Howard1} and \cite{Huang}. The results obtained were qualitatively
the same as those shown above. This indicates that the observed behaviour
is not a result of the stochastic nature of our model.


\section{Conclusion and Outlook}

In this paper we have introduced a new model for the Min protein
oscillations, incorporating both membrane polymerisation and stochasticity.
As we have seen, the model is able to account for much of the observed 
Min dynamics. We have also applied our model to the dynamics of the 
Min proteins during cell division and found that diffusion alone is
insufficient to equalise the protein copy numbers between 
the two daughter cells.

Although there have been a few comments on Min dynamics in constricting cells,
there have been no experiments looking systematically and quantitatively
at protein dynamics in large numbers of cells undergoing the division process.
We hope that future experiments will investigate the partitioning of the
Min proteins and follow the Min oscillations into the daughter cells. Although
the results we have presented appear to be general and independent of the
division mechanism, it is possible that other models would produce different
behaviour. This provides potentially another way to test these models against
experimental observations and each other.

The reports of oscillations in constricting cells \cite{RdB,Hu1} have
stated that oscillations of the Min proteins continue unaffected well
into the division process. After this time, oscillations occur
separately between each pole and mid-cell, and continue once the
daughter cells have separated. These features are reproduced in our
simulations - oscillations cut off sharply at some time during the
closing of the septum, after which the daughter cells are effectively
independent even though they have not yet completely separated.
 
Our simulations suggest that the distribution of the Min proteins is
very often unequal and often largely skewed to one daughter cell. The
variation of periods observed \textit{in vivo} also leads us to
believe that there is some variation of copy number between cells.
However, in the most extreme cases of our simulations, Min
oscillations are not supported in the daughter cells. Wild-type
\textit{E. coli} without pole-to-pole Min oscillations have not been
reported in the literature. It may be that our model cannot reproduce
oscillations at the extremes of the range where they can occur
\textit{in vivo}. However, in these cases the period of oscillation
would probably lie well outside the range typically observed. This
suggests that, at least in these extreme cases, some additional way of
regulating protein numbers in the daughter cells may be required.

For most cytoplasmic proteins that are present in high numbers, diffusion
effectively distributes them evenly throughout the cell so that at division
the number in each daughter cell is roughly equal. The dynamics of the Min
proteins, however, means that the distributions are normally skewed greatly
towards one end of the cell. From our simulations we conclude that diffusion
through the septum is not by itself able to equalise the Min protein numbers
in each daughter cell.

The final expression levels in the daughter cells appear to be entirely uncorrelated
with the fractions of proteins in each half of the parent cell at the beginning
of the division process. This is presumably a result of the inherently stochastic
nature of the model dynamics. This means it is unlikely
that the initiation of division can be controlled to coincide with a certain
point in the Min cycle to ensure equal $\rho_{\rm D}:\rho_{\rm E}$ ratios
in the daughter cells. An alternative would be some form of active transport
through the closing septum. This also appears unlikely, and there is certainly
no experimental evidence for such a mechanism. It therefore seems improbable
that the protein numbers are regulated by the division mechanism itself,
which leaves open the possibility that levels are corrected shortly after
division.

In our simulations, those cells which did not have Min oscillations had a
$\rho_{\rm D}:\rho_{\rm E}$ ratio below 1.6. This could be rectified by producing
more MinD shortly after division. Additionally those cells with a very low
copy number of both proteins had small and low-density polar zones, where
fluctuations had a much more significant impact on the pole-to-pole oscillations,
leading to a much less pronounced MinD mid-cell concentration minimum. These
cells would also benefit from increased copy numbers of both proteins. This
could be achieved if the production rate of the Min proteins is controlled
according to their concentrations, without needing a direct trigger from
the division event. The production of the Min proteins has yet to be studied
experimentally, so it is not known which, if any, factors affect their production
rates.

Previous studies \cite{rueda,arends} have found that there is no
evidence for cell-cycle dependent protein synthesis in
\textit{E. coli}, including cell division proteins such as FtsZ and
FtsA. For proteins involved in the division machinery such as FtsZ, a
constant production rate is sufficient for these proteins to be
equally distributed at cell division. The majority of FtsZ is
cytoplasmic and so the concentration throughout the parent cell would
be largely equalised by diffusion. The remaining FtsZ is located at
the septum in the ``Z-ring", and proteins in this structure could
easily be equally divided between the daughter cells. 

However, as described above, the situation for the Min proteins is likely
to be rather different. Potentially the concentration levels of the Min proteins
may feedback to their production (or even degradation) rates, so that, for
example, their rates of synthesis increase whenever their concentrations
are low. After division some cells would therefore have a burst of protein
synthesis, but this would not be directly triggered due to the cell having
recently divided. As the cell continues to grow the same mechanism could
also keep the Min protein concentrations roughly constant. In future experiments
it will be interesting to thoroughly test some of these possibilities.

\ack We thank J. Lutkenhaus for a useful discussion. F.T. is
supported by the EPSRC, and M.H. by The Royal Society.

\section*{Glossary}
{\it ATP hydrolysis:} Highly exothermic chemical reaction which is a major
source of energy inside a cell. MinE stimulates the hydrolysis of ATP associated
with polymerised MinD, which leads to polymer disassembly. \\
{\it Cell cycle:} The sequence of events a cell goes through in order to
reproduce, including DNA replication and division.
\\
{\it Pattern formation:} Appearance in an extended system of spatially- and/or
time-varying structures. \\
{\it Reaction-diffusion system:} A system of particles, each of which moves
by diffusion, and where reactions occur when particles meet. These systems
can sometimes display pattern formation.
\\
{\it Filamentous cell:} A cell which is prevented from dividing, usually
by removal of FtsZ, and which therefore grows much longer than is typical
in wild-type bacteria.

\section*{References}

\bibliography{min}

\end{document}